\documentclass[journal]{IEEEtran}
\usepackage{cite}
\usepackage{amsmath,amssymb,amsfonts,bbold}
\usepackage{bm}
\usepackage{algorithm2e}
\usepackage{enumitem}
\usepackage{tabularx}
\usepackage{makecell}
\usepackage{graphicx}
\usepackage{textcomp}
\usepackage{multirow}
\usepackage{dblfloatfix}
\usepackage{lipsum}
\usepackage{caption}
\usepackage{balance}
\ifCLASSOPTIONcompsoc
 \usepackage[caption=false,font=normalsize,labelfont=sf,textfont=sf]{subfig}
\else
 \usepackage[caption=false,font=footnotesize]{subfig}
\fi

\usepackage{hyperref}
\hypersetup{colorlinks,allcolors=black}

\newcolumntype{L}[1]{>{\raggedright\let\newline\\\arraybackslash\hspace{0pt}}p{#1}}

\begin{document}
\title{Achieving AI-enabled Robust End-to-End Quality of Experience over Radio Access Networks

% AI-enabled Robust Orchestration of \\ Network and Computing Resources \\for End-to-End Quality of Experience 
%Slicing in O-RAN and Edge-computing for End-to-End Quality of Experience %
}
%%%%%%%%%%%%%%%%%%%%
\author{
Dibbendu~Roy, 
Aravinda~S.~Rao,~\IEEEmembership{Senior~Member,~IEEE,}
Tansu~Alpcan,~\IEEEmembership{Senior~Member,~IEEE,} \\
Goutam Das, and
Marimuthu Palaniswami,~\IEEEmembership{Fellow,~IEEE}
\thanks{This research was supported partially by the Australian Government through the Australian Research Council's Discovery Projects funding scheme (project DP190102828).}
\thanks{D. Roy, A. S. Rao, T. Alpcan and M. Palaniswami are with the Department of Electrical and Electronic Engineering, The University of Melbourne, Parkville, Victoria - 3010, Australia (e-mail: dibbendu.roy@student.unimelb.edu.au, aravinda.rao@unimelb.edu.au, tansu.alpcan@unimelb.edu.au, palani@unimelb.edu.au). }
\thanks{G. Das is 
now with the G.S. Sanyal School of Telecommunications, Indian Institute of Technology Kharagpur, West Bengal - 721302, India. (e-mail: gdas@gssst.iitkgp.ac.in).}
}
%%%%%%%%%%%%%%%%%%%%

\markboth{IEEE Transactions on Cognitive Communications and Networking ,~Vol.~0, No.~0, 
November~2021}%
{Roy \MakeLowercase{\textit{et al.}}: A Sample Article Using IEEEtran.cls for IEEE Journals}

\maketitle

\begin{abstract}
Emerging applications such as Augmented Reality, the Internet of Vehicles and Remote Surgery require both computing and networking functions working in harmony. The End-to-end (E2E) quality of experience (QoE) for these applications depends on the synchronous allocation of networking and computing resources. However, the relationship between the resources and the E2E QoE outcomes is typically stochastic and non-linear.  In order to make efficient resource allocation decisions, it is essential to model these relationships. This article presents a novel machine-learning based approach to learn these relationships and concurrently orchestrate both resources for this purpose. The machine learning models further help make robust allocation decisions regarding stochastic variations and simplify robust optimization to a conventional constrained optimization. When resources are insufficient to accommodate all application requirements, our framework supports executing some of the applications with minimal degradation (graceful degradation) of E2E QoE. We also show how we can implement the learning and optimization methods in a distributed fashion by the Software-Defined Network (SDN) and Kubernetes technologies. Our results show that deep learning-based modelling achieves E2E QoE with approximately 99.8\% accuracy, and our robust joint-optimization technique allocates resources efficiently when compared to existing differential services alternatives.
\end{abstract}


\section{Introduction}\label{sec:introduction}
New applications and use cases are largely stimulated by modern networking architectures and solutions for 5G and beyond. The International Telecommunication Union has classified 5G mobile network services into three categories: (1) Enhanced Mobile Broadband (bandwidth-intensive like Virtual Reality), (2) Ultra-reliable and Low-latency Communications (highly delay-sensitive and (3) reliable like automated driving), and Massive Machine Type Communications (high connection density like IoT and Industry 4.0) \cite{sachs20185g}. In contrast, in 6G, there might be further granularity involved in defining these categories \cite{letaief2019roadmap} based on applications. In addition, 6G aims at integrating intelligence in the network, implying that the network should interpret user requirements and take suitable decisions to satisfy them. For example, Australian telecommunication service provider Telstra promises at delivering programmable networks as per user needs \cite{telstra}.

User requirements are typically built upon classical metrics such as delay, throughput, jitter, and reliability. The aforementioned new applications require satisfying end-to-end (E2E) Quality of Experience (QoE) based on these metrics. However, the relationship between the E2E QoE parameters and the resources is non-linear (typically non-convex) and stochastic \cite{she2019cross}. It also challenges service providers to allocate resources to satisfy the E2E QoE requirements of emerging applications. Most of the available works either assume that the relationship between E2E QoE and resources is simplistic or that the users are aware of the required resources for achieving the desired E2E QoE. However, these assumptions might not be realistic given the complex and dynamic nature of these applications.

Even though we talk of networks, these applications require services from both networking and computing. 
Modern networking technologies such as \textit{software defined networks (SDN)} \cite{chartsias2017sdn} and and computing technologies like \textit{Kubernets (K8)}\cite{kubernetes_2021} and \textit{dockers}\cite{docker} enable virtualization of networks and applications, which allows each application to have its own network and computing resources (refer Section \ref{sec:litrev}). Thus, service providers must jointly design and adapt their networking and computing resources to satisfy the diverse E2E QoE requirements. Although a plethora of research work is available for satisfying QoE metrics separately in network and computing domains, a comprehensive study on the interaction of the two while considering the relevant technologies in these paradigms is missing (refer Section \ref{sec:litrev}).  

This paper presents a modern approach to joint computing and networking resource orchestration. Building upon the technological aspects involved in characterizing E2E QoE, we develop a sequential machine learning and optimization-based framework to meet E2E QoE requirements, primarily focusing on robustness aspects.

The contributions of the paper can be summarized as:
\begin{itemize}
    \item We address the problem of E2E QoE satisfaction in emerging applications with the help of modern networking and computing technologies. To the best of our knowledge, this is the first work that combines both networking and computing aspects concerning E2E QoE while considering SDN and Kubernetes as orchestrators of the network and computing sites.
    \item We consider the fronthaul segment of a radio access network with bursty traffic arrivals. The relationship between the resources and the E2E QoE parameters will depend on several parameters, some of which may also have stochastic variations. In this paper, we propose to use AI to model these relationships with the help of deep learning. We develop a novel framework based on sequential learning and optimization for E2E QoE satisfaction. The learning models helps to simplify a robust optimization problem to a conventional constrained optimization by directly learning the worst case possibilities. 
    We show how our approach can be implemented as a digital twin. \cite{Cheatshe79:online} using emulation platform.
    \item When network or computing resources are constrained, it is desirable to have minimal service degradation depending on service level agreements. In these situations, services can be optimally reconfigured or gracefully degraded with the help of the presented optimization framework. 
    \item Implementing a digital twin might not be an efficient solution to many service providers as it comes at cost of additional processing and security threat. The proposed approach can be applied on a real network setting if SDN conroller and Kubernetes can work indepedently with minimal coordination. Thus, it is important to learn and optimize in a distributed manner. We present the challenges in the distributed learning scenario and show how the problem can be tackled with help of primal decomposition technique. This implementation comes at cost of additional storage for training over all examples rather than the worst ones. 
    \item Our results show the efficacy of using deep learning towards achieving uRLLC requirements. Compared to existing methods, it can achieve uRLLC targets of reliability of more than 99\%. Further, we show the effect of using joint-optimization instead of the existing prioritization standards at network and computing sites. While joint-optimization can satisfy the uRLLC requirements, the same is not the case for other approaches. We also demonstrate the utility of graceful degradation through our approach.
\end{itemize}
 
The rest of the paper is organized as follows: Section \ref{sec:litrev} contains relevant background and gaps in literature leading to Section \ref{sec:probmetmod} which discusses the problem, model and methodology for our learning and optimization approach. We discuss the learning methodology followed by joint robust optimization techniques in section \ref{sec:learn} and \ref{sec:joint}. Finally, we present the relevant emulation and simulation results in Section \ref{sec:expresults} followed by concluding remarks in Section \ref{sec:sconclusion}.

\section{Background and Literature Review} \label{sec:litrev}

Modern networking technologies move away from traditional solutions, which require specialized networking hardware and are challenging to maintain and configure. It is envisioned that 6G networks will be developed over SDN \cite{chartsias2017sdn} which separate the control and data planes at the network devices. The notion of separation simplifies hardware requirements for network devices and increases flexibility as we can program them from a centralized controller. We can implement networking operations such as routing, bandwidth allocation, and security using virtual machines (VMs) on a general server which is commonly known as \textit{network function virtualization (NFV)}  \cite{chartsias2017sdn}. From a purely computing perspective, there have been some critical advances in virtualization technologies as well. The concept of having VMs with hardware-level segregation of resources has been replaced with lightweight middleware applications called \textit{containers} \cite{docker}. Similar to VM, each container has its own set of resources managed at the OS level with the help of software such as \textit{dockers}. For container deployments over multiple servers, production-level container orchestration software, such as \textit{Kubernetes (K8)} \cite{kubernetes}, manages and maintains the containers. With the help of these emerging technologies, service providers can satisfy E2E QoE requirements by implementing a logical allocation of the network and computing infrastructure, also known as \textit{network slicing}. Based on these technologies, next-generation access networks are moving towards Open Radio Access Network (O-RAN) \cite{niknam2020intelligent}, which aims to provide RAN functionalities on open hardware and software solutions. 

We now provide a brief review of relevant research literature in this area. There are several proposals for resource allocation in the case of networks and cloud computing for achieving QoE \cite{shu2020novel,dutra2017ensuring,oliveira2018sdn,sattar2019optimal,dietrich2017network,chartsias2017sdn,ordonez2017network,8082092,8247221,baumgartner2017optimisation,jiang2017network,rossi2020hierarchical,hirai2020automated,wiranata2020automation,figueiredo2020edgevpn,gawel2019analysis,wei2020network}. Broadly, we can classify these works as those involving computing (Kubernetes based) \cite{rossi2020hierarchical,hirai2020automated,wiranata2020automation,figueiredo2020edgevpn,gawel2019analysis} and the ones involving networking (SDN based) \cite{dietrich2017network,shu2020novel,dutra2017ensuring,oliveira2018sdn,sattar2019optimal,chartsias2017sdn,ordonez2017network,8082092,8247221,baumgartner2017optimisation,jiang2017network}. The slicing approaches formulated over SDN typically find/reconfigure routes and decide on bandwidth allocations along the routes. The authors of \cite{shu2020novel,dutra2017ensuring} evaluate their performance by calculating the time taken to create a network slice. Some existing works are listed in \cite{wei2020network} which mainly focus on re-configuring slices using AI-based techniques. These works adapt to changing slice requirements.  However, while designing their slice configuration or re-configuration algorithms, they assume that the applications know the resource requirements. This assumption is quite restrictive as applications or users would be only concerned about E2E QoE, and the mapping between E2E QoE and resources is often complex, dynamic, and stochastic. 

The authors in \cite{rossi2020hierarchical,hirai2020automated,wiranata2020automation,figueiredo2020edgevpn,gawel2019analysis} mostly deal with Kubernetes settings and container deployment strategies. However, they do not incorporate the associated networking delays. Thus, E2E QoE has not been investigated in these works. 

Although most of the papers mentioned above do not consider the roles of SDN and Kubernetes in their models, \cite{sattar2019optimal,ordonez2017network} consider network, computing and E2E QoE. In these works, the authors solve the problem of VM placement based on server and network loads which turns out to be a Mixed-Integer linear optimization problem (MILP). The authors consider a simple linear relationship of E2E delays concerning network loads which is not the case in reality due to randomness and non-linear behaviour of delay concerning loads, as evident from elementary queuing theory. Further, they do not consider the effect on E2E delays due to the resulting placement strategies.

\section{Problem, Model, and Methodology}
\label{sec:probmetmod}
\subsection{Problem Statement}
From our previous discussions in Sections \ref{sec:introduction} and \ref{sec:litrev}, it is clear that there have been very few works that consider E2E delays from both networks and computing perspectives. In existing works, applications/tasks have their bandwidth and processing requirements, and they place these at servers where they can achieve load balancing. Thus, the papers mentioned above have modeled these problems as assignment problems requiring binary decisions resulting in mixed-integer problems. Apart from the assignment, modern computing technologies such as Dockers and Kubernetes allow deploying applications with different resources. The decision implies that the requisite computing resources must be assigned to ensure E2E QoE. Users should only care about their experience while the resource allocation decisions (networking and computing) are intelligently decided as envisioned in 6G networks. This paper proposes techniques to model E2E QoE requirements and subsequently perform joint multi-resource allocation of networking and computing resources to satisfy the desired E2E QoE requirements.

In this paper, we address the following problems:
\begin{itemize}
    \item How to model E2E QoE with respect to bandwidth and processing as resource variables?
    \item How to obtain robust optimal decisions regarding resource variables? 
    \item How to handle resource constrained situations?
    \item How to implement our approach as a digital twin using emulation environments?
    \item How to implement the allocation decisions with independent network and computing orchestrators?  
\end{itemize}

Before we proceed to our methodology, we present the relevant notations and system model for our work.

\subsection{Model and Notation}
\begin{figure}
    \centering
    \includegraphics[height=5cm]{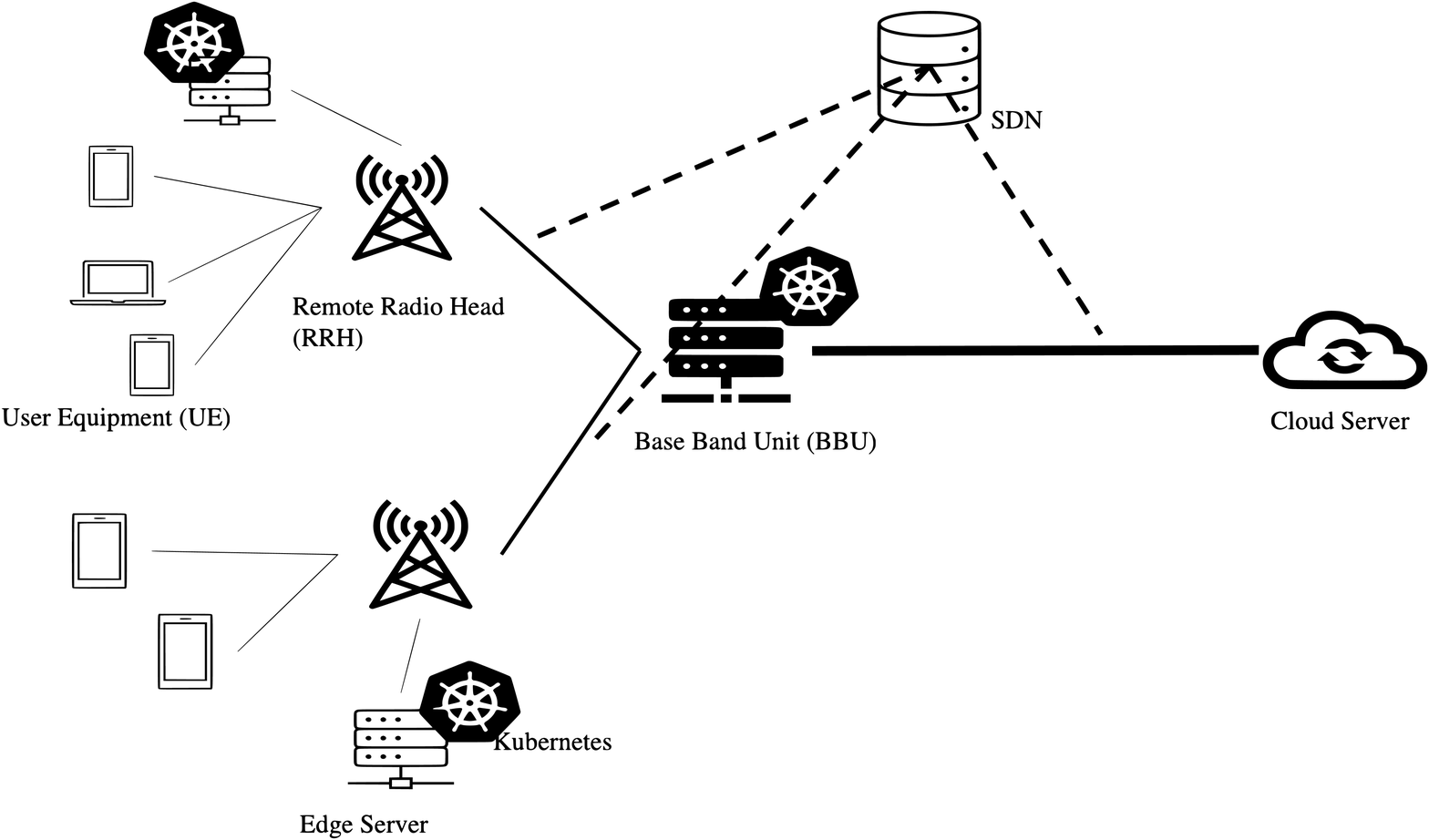}
    \caption{A 5G Radio Access Network}
    \label{fig:RAN}
\end{figure}
We consider a Radio Access Network (RAN) where users are connected to a Baseband Unit (BBU) through a remote radio head (RRH) and a router (see Fig. \ref{fig:RAN}). The BBU is equipped with Kubernetes or Docker system to manage containers (applications) that run on them. An SDN controller can configure the router, and users host different applications with QoE requirements served at the BBU. The SDN and Kubernetes create E2E slices. They can communicate and share statistics related to QoE metrics required to decide the resources for a slice. 

Users host different applications or classes, each with different QoE requirements. A slice comprises applications having similar QoE requirements. In our model, we consider that an application sends packets related to the application to be executed at an edge or cloud server. The packets carry the required data for executing the job. For example, an application might be an algorithm to sort an array where packets contain array data. In such applications, the type of application and the amount of data it sends determine the degree of processing required at the server. Thus, each application can have its traffic generation rates and processing demands.

Further, depending on the type of protocol employed at MAC and Transport layers, the delays experienced at the network sites may vary. An E2E user experience can thus be characterized by the delays experienced to obtain the result of an application request. Also, there might be situations when packets get lost either at the network (considering a UDP based protocol for real-time applications) or at the server (due to overload or out of memory). In such cases, the user does not get a response from the server leading to an undesirable experience. To model this effect, we define E2E throughput as the percentage of successful requests. Thus, we use the two metrics, namely E2E delay and E2E throughput, for characterizing user experience.

Let $\mathbb{A} \in \{a_1,\dots,a_N\}$ denote the set of application classes or slices with different QoE requirements, requested by the users. Each class $a_i$ demands a QoE depending on its requirements given by the tuple 
$q_i = (\tau_i,\rho_i)$, where $\tau_i$ denotes the desired E2E delay, $\rho_i$ denotes the desired E2E throughput. E2E delay comprises of all relevant delays such as access, propagation, queuing, transmission and processing in the network and the computing sites. 

% E2E throughput can be defined as the percentage of successful requests that could be handled by the network and server for an application class $a_i$.

\begin{table}[!t]
\renewcommand{\arraystretch}{1.3}
\caption{Notations used in this work.}
\label{notation_tab}
\centering
\begin{tabular}{|p{1.5cm}|p{6cm}|}
    \hline
    \multicolumn{1}{|c|}{\textbf{Notation}} & \multicolumn{1}{|c|}{\textbf{Description}}  \\
    \hline
    $\mathbb{A}$ & Set of Applications\\
    \hline
    $G=(V,E)$ & Physical network graph. $V$ is set of vertices and $E$ is the set of edges/links. Routing decisions rely on link capacities.\\
    \hline
    $B(e)$ & Capacity of link $e \in E$\\
    \hline
    $\theta_{i}$ & Random variable to model uncertainty in E2E QoE due to stochastic variations \\
    \hline
    $\theta_{i}$ & Range Space of $\theta_{i}$ i.e., $\theta_{i} \in \theta_{i}$\\
    \hline
    \multicolumn{2}{c}{\textbf{QoE Parameters}}\\
    \hline
    ${\tau_i}$ & E2E delay requirement for application $a_i\in \mathbb{A}$\\
    \hline
    $\rho_i^N$ & Throughput requirement from network for application $a_i\in \mathbb{A}$\\
    \hline
    $\rho_i^S$ & Throughput requirement from server for application $a_i\in \mathbb{A}$\\
    \hline
    $r_i$ & Probability figure requirement that $a_i\in \mathbb{A}$ does not fail either due to network or server problems\\
    \hline
    \multicolumn{2}{c}{\textbf{Slicing Decision Variables}}\\
    % \hline
    % $b_i$ & Bandwidth allocation for application at each link (implemented by SDN/Network Controller)\\
    \hline
    $f_i^e$ & Flow for $a_i\in \mathbb{A}$ along edge $e\in E$ (implemented by SDN/Network Controller)\\
    \hline
    $\phi_i^c$ & Fraction of CPU core $c$ allocated at the server side  for $a_i\in \mathbb{A}$ (implemented by Kubernetes)\\
    \hline
    \multicolumn{2}{c}{\textbf{Functions in Optimization}}\\
    \hline
    $D_i()$ & Delay constraint function\\
    \hline
    $T_i()$ & Throughput constraint function \\
    \hline
    $u_i()$ & Utility function (can be average delay, sum throughput, cost etc.)\\
    \hline
    $\psi_i()$ & Penalty function for graceful degradation of constraint to be relaxed\\
    \hline
\end{tabular}

\end{table}

Typically, network slicing involves bandwidth slice, path/route and the corresponding processor allocation at the edge/cloud server. Given a graph $G = (V,E)$ and a server with $C$ cores, a slicing decision may be modeled as a tuple $ (f_i^e,\phi_i^c)$ $\forall e\in E$ and $\forall c \in C$, where $f_i^e \in [0,1]$ denotes the flow rate for $a_i$ along edge $e$. This decision on $f_i^e$ is taken by an SDN controller or network hypervisor like FlowVisor \cite{GitHubop19:online}. $\phi_i^c \in [0,1]$ denotes the fraction of the processing for core $c$ at a server. This modeling allows us to capture parallelizable applications as well, and its realistic implementations are feasible due to the advent of dockers and Kubernetes based server systems \cite{docker,kubernetes_2021}. Like CPU allocation decisions, Kubernetes can also decide on the amount of RAM to be allocated for a slice. This can also be introduced as a decision variable along with $\phi_i^c$. However, we found that changing the amounts of RAM do not significantly affect the E2E delays and E2E throughput. In our experiments and this paper, we assume that the server has sufficient RAM to execute an application. Thus, we have ignored the decision variable for RAM in the paper. However, there might be situations where RAM might be important while handling large matrices or such operations. In such cases, the model can be suitably extended. The variables for each $e$ and $c$ may be collected and stacked as a vector. We denote this by bold letters. However, for the considered RAN network, the decisions boil down to scalar variables due to single link and consideration of non-parallel applications in the paper.

Each of the two major E2E QoE parameters, namely delay and throughput can be captured via implicit functions of the form:
\begin{align}
\text{E2E Delay: } & D_i(\bm{f_i^e},\bm{\phi_i^c};\bm{\theta_{i}})\\
\text{E2E Throughput: }&T_i(\bm{f_i^e},\bm{\phi_i^c};\bm{\theta_{i}})
\end{align}

Here, E2E throughput is determined by the number of successful requests. A request is successful if it is successfully transferred by the network and subsequently processed at the server. Other QoE parameters can also be included to extend our model. Although we write implicit functions for mapping the QoE constraints to decision variables, it is essential to note that these functions may also depend on one or more stochastic parameters. These parameters are denoted by $\bm{\theta_{i}}$ which can be associated with a range space $\bm{\Theta_{i}}$. A simple example of such a parameter is the number of customers/subscribers for a given slice. Since in this paper, we consider a RAN network, the vector parameters are ignored henceforth and we use scalar notations.

\subsection{Methodology}

% \begin{figure}[h]
%     \centering
%     \includegraphics[width=0.95\linewidth]{Figures/flow-slicing_v2.eps}
%     \caption{Proposed approach for network slicing with learning and optimization phases.}
%     \label{fig:algo}
% \end{figure}

As discussed in Section \ref{sec:introduction}, this paper focuses on satisfying QoE requirements over the 5G network and beyond. The QoE performances depend on allocated computing and networking resources. It is well understood that the resource allocation with QoE constraints can be cast as a constrained optimization problem as has been primarily explored in the literature (refer Section \ref{sec:litrev}). However, the QoE parameters may exhibit complex dependencies with those of the resources, and hence such problems have been difficult to solve in general \cite{8247221,8082092,baumgartner2017optimisation}. 

\begin{figure*}[!t]
    \centering
        \includegraphics[width=0.95\linewidth]{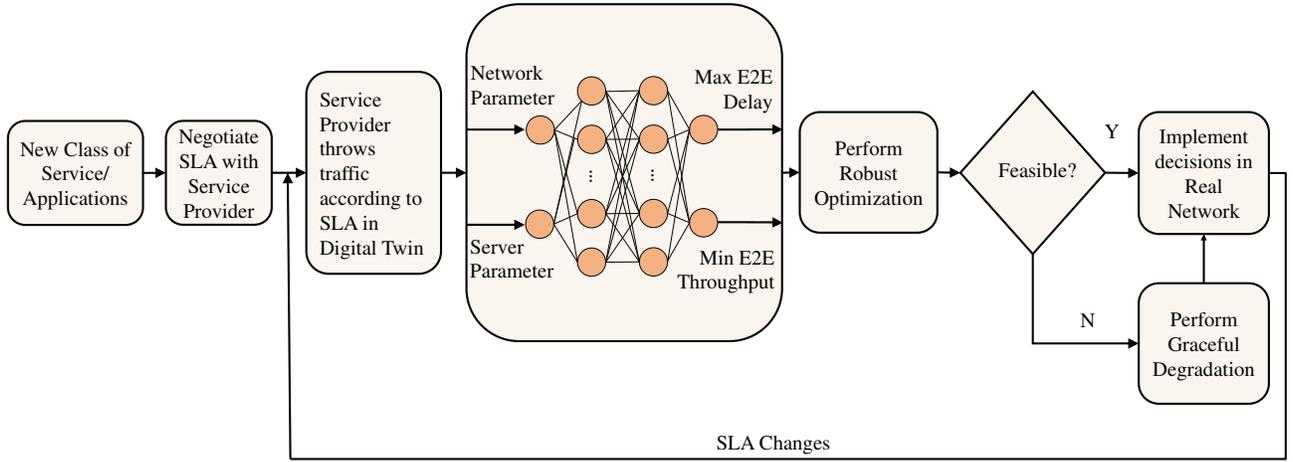}
    \caption{Conceptual View of Sequential Learning and Joint Multi-Resource Allocation for E2E QoE using Digital Twin}
    \label{fig:concept}
\end{figure*}

In this work, we try to create network slices in two phases. In the first phase, we exploit the power of deep learning to learn these complex dependencies. In the second phase, we solve an optimization problem to obtain allocation/slicing decisions. When we say a slice, it implies creating a network slice and corresponding allocation at the server-side. Our approach for solving the problem can be depicted as shown in Fig. \ref{fig:concept}. When a service provider receives a new slice request (QoE requirements), a service provider may implement the digital twin with help of available emulation platforms for SDN and Kubernetes (refer Section \ref{res:emu}). It then probes the network and server according to agreed-upon service level agreements. By varying the networking and computing resources, one can obtain the desired QoE metrics. A deep neural network is very efficient in learning any arbitrary function \cite{kratsios2021universal}. Fig. \ref{fig:concept} shows a conceptual overview of how our approach may be implemented in a digital twin \cite{Cheatshe79:online}.  Using the digital twin, the E2E delays and E2E throughput of clients may be emulated for various application requests. Then a neural network is trained to capture the complex relationship between the resource parameters and QoE metrics.

% The SDN controller at the network site and Kubernetes at the server site can employ individual neural networks to learn and optimize in a distributed manner, as shown in Fig. \ref{fig:concept}.
Once these dependencies are obtained, one can solve an optimization problem to obtain robust network slices. In case that resources are not enough to satisfy the QoE requirements, some slices might have to be reconfigured so that critical slices satisfy their QoE requirements at the cost of relaxed QoE satisfaction of other slices. We term this as graceful degradation of services which is desirable in the case of resource-constrained scenarios. In this work, we assume that the service level agreements are negotiated between the slices and service providers, allowing the service providers to probe network and servers suitably. These decisions might change with newly negotiated or renegotiated agreements. The loop in our proposed approach Fig. \ref{fig:concept} captures this effect.
% \section{Learning the E2E QoE Model}
% \label{sec:learn}
% % \subsection{E2E QoE Constraint Functions}
% % \label{QoEcon}
% %  Ideally the slicing strategy should be optimal even for the \textit{worst} parameter settings which typically falls under the category of robust optimization \cite{bertsimas2011theory}.  
% % In the later scenario, the problem falls under the paradigm of stochastic optimization \cite{birge2011introduction}. In this paper, we consider the robust counterpart.
% % \subsection{Learning Constraint Functions}
% It is well known that by Universal Approximation Theorem \cite{kratsios2021universal}, neural networks act as universal function approximators \cite{goodfellow2016deep} and hence can be used to learn or approximate arbitrary dependencies (see Fig. \ref{fig:concept}). Based on the nature of the constraints, one may apply a suitable neural network to approximate its nature.

% \begin{figure}
%     \centering
%     \includegraphics[scale=0.35]{Figures/nn.eps}
%     \caption{Distributed Learning of E2E QoE Dependencies}
%     \label{fig:neural}
% \end{figure}

% \textbf{ToDo: The section below is a bit confusing. You should clearly focus on what we are learning and how. How do we collect data? (3) and (4) are nice, but do you train separate NNs for each component in them?}

%  

\section{Sequential Learning and Joint Multi-Resource Allocation}
\label{sec:joint}
We adopt learning based robust optimisation approach for multi-resource allocation. Specifically, we present the constraints, objective, and robust optimization approach for joint allocation of network and computing resources. Apart from the E2E QoE constraints, the capacity constraints for communication links and processing of the server should be taken into consideration for joint optimization.
% \section{Learning E2E QoE dependencies}

\subsection{E2E QoE Constraints}
% \label{QoEcon}
% Each of the E2E QoE parameters  - delay and throughput may be imposed by considering implicit functions of the form:
The QoE functions discussed before are restricted by the following bounds:
\begin{align}
D_i(f_i^e,\phi_i^c;\mathbf{\theta_{i}})&\leq \tau_i ~~~\forall ~i,\theta_{i} \label{eq:delcon}\\
T_i(f_i^e,\phi_i^c;\mathbf{\theta_{i}}) &\geq \rho_i ~~~\forall ~i,\theta_{i} \label{eq:throughcon}
\end{align}

Here, $D_i(f_i^e,\phi_i^c;\mathbf{\theta_{i}})$ and $T_i(f_i^e,\phi_i^c;\mathbf{\theta_{i}})$
are the learned functions discussed in Section \ref{sec:learn}
\subsection{Learning the E2E QoE Model}
\label{sec:learn}
% \subsection{E2E QoE Constraint Functions}
% \label{QoEcon}
%  Ideally the slicing strategy should be optimal even for the \textit{worst} parameter settings which typically falls under the category of robust optimization \cite{bertsimas2011theory}.  
% In the later scenario, the problem falls under the paradigm of stochastic optimization \cite{birge2011introduction}. In this paper, we consider the robust counterpart.
% \subsection{Learning Constraint Functions}
It is well known that by Universal Approximation Theorem \cite{kratsios2021universal}, neural networks act as universal function approximators \cite{goodfellow2016deep} and hence can be used to learn or approximate arbitrary dependencies (see Fig. \ref{fig:concept}). Based on the nature of the constraints, one may apply a suitable neural network to approximate its nature. To account for the stochastic nature of network state and server, service provider uses a Monte-Carlo strategy (by using different random seeds to generate traffic) and obtain the required statistics. For example, for a given bandwidth and processing configuration, one can obtain the histogram for E2E delays experienced by the packets. Once the data regarding the QoE metrics are collected, a neural network is trained as shown in Fig. \ref{fig:concept}, for a digital twin based implementation.
\subsection{Link Capacity}
Given a link $e\in E$ from the network graph $G$, the capacity of the link is bounded by $B(e)$. The bandwidth allocated to applications routed through this link should not exceed this capacity. Since $f_i^e$ denotes the probability that application $a_i$ is routed through the link $e$, we have

\begin{equation}
\sum_{i|a_i\in \mathbb{A}} f_i^e \leq 1 ~ ~~\forall e \in G
    % \implies \sum_{i|a_i\in \mathbb{A}} f_i^e \leq 1 ~ ~~\forall e \in G
\label{eq:linkcap}
\end{equation}

% 
% \textbf{ToDo: this is simply wrong if f is a probability.}

\subsection{Server Capacity}
The fraction of processing for a given core $c$ at a server should not exceed the maximum processing capability of the core. We consider a core as a unit processing element and all cores are identical. Thus, we must have
\begin{equation}
    \sum_{i|a_i \in \mathbb{A}} \phi_i^c \leq 1 ~~~\forall ~c \in C
    \label{eq:servcap}
\end{equation}

Here, $C$ denotes the set of cores of a CPU in a server.

\subsection{Objective Function} 
Network Slicing can be performed depending on many desired objectives such as sum throughput maximization, delay minimization, minimization of delay violation, load balancing etc. Without loss of any generality, we consider general utility functions which may be mapped to any of the aforementioned objectives (similar to well-known network utility maximization). Let $u_i(f_i^e,\phi_i^c)$ denote the utility obtained on executing application $a_i$ over an edge network. Then, a generic objective might be of the form:
\begin{equation}
    \sum_{i|a_i\in \mathbb{A}} u_i(f_i^e,\phi_i^c;\theta_{i})
    \label{eq:utility}
\end{equation}
% \textbf{ToDo: say specifically what we use in this paper as examples}
In this paper, we use throughput as the utility. $u_i(f_i^e,\phi_i^c;\theta_{i}) = T_i(f_i^e,\phi_i^c;\theta_{i})$
% \subsection{Optimization Techniques}
% We present all generalizations of the network slicing problem. 
% \subsubsection{Network Slicing optimization problem}
% We start with the simple case where we do not consider parameter variations.
% Then, the network slicing problem may be devised as:
% \begin{subequations}
% \label{slicingopt}

% \begin{equation}
%     \max_{(f_i^e,\phi_i^c)} \sum_{i|a_i\in \mathbb{A}} u_i(f_i^e,\phi_i^c)
% \end{equation}
% \begin{equation}
% \text{s.t.} ~ D_i(f_i^e,\phi_i^c)\leq \tau_i ~~~\forall ~i
% \end{equation}
% \begin{equation}
% T_i^N(f_i^e,\phi_i^c) \geq \rho_i^N ~~~\foralsl ~i
% \end{equation}
% \begin{equation}
% T_i^S(f_i^e,\phi_i^c) \geq \rho_i^S ~~~\forall ~i
% \end{equation}
% \begin{equation}
% R_i(f_i^e,\phi_i^c)\geq r_i ~~~\forall ~i
% \end{equation}
% \begin{equation}
% \sum_{i|a_i\in \mathbb{A}} f_i^e \leq B(e) ~ ~~\forall e \in G
% \end{equation}
% \begin{equation}
%     \sum_{i|a_i \in \mathbb{A}} \phi_i^c \leq 1 ~~~\forall ~c \in C
% \end{equation}
% \end{subequations}

\subsection{Robust Optimization using Machine Learning}
Before dealing with the distributed version of the problem, we state the robust centralized version of the problem. This can be directly implemented in a digital twin based implementation. A robust design is one which works for all possible variations of the parameters in the parameter space. Thus, by robustness we strive to design slices with worst possible parameter combinations. Hence, we may pose the robust optimization problem as a max-min problem, where we maximize our objective subject to worst possible effect of the parameters, in this case $\theta_{i}$ on the problem \cite{bertsimas2011theory}. 

Using \eqref{eq:delcon},\eqref{eq:throughcon},\eqref{eq:linkcap},\eqref{eq:servcap} and \eqref{eq:utility}, we may state the robust Optimization problem as: 
% \textbf{ToDo: say now that this is the combined problem bringing introduced elements together}
\begin{subequations}
\label{slicingopt_robust}
\begin{equation}
    \max_{(f_i^e,\phi_i^c)} \min_{\theta_{i} \in \theta_{i}}\sum_{i|a_i\in \mathbb{A}} u_i(f_i^e,\phi_i^c;\theta_{i})
    \label{fitness}
\end{equation}
\begin{equation}
\text{s.t.} ~ D_i(f_i^e,\phi_i^c;\mathbf{\theta_{i}})\leq \tau_i ~~~\forall i,\theta_{i} 
\label{con1}
\end{equation}
\begin{equation}
T_i(f_i^e,\phi_i^c;{\theta_{i}}) \geq \rho_i ~~\forall i,\theta_{i}
% T_i^N(f_i^e,\phi_i^c;\mathbf{\theta_{i}}) \geq \rho_i^N ~~~\forall ~i
\label{con2}
\end{equation}
% \begin{equation}
% T_i^S(f_i^e,\phi_i^c;\mathbf{\theta_{i}}) \geq \rho_i^S ~~~\forall ~i
% \end{equation}
% \begin{equation}
% R_i(f_i^e,\phi_i^c;\mathbf{\theta_{i}})\geq r_i ~~~\forall ~i
% \end{equation}
\begin{equation}
\sum_{i|a_i\in \mathbb{A}} f_i^e \leq 1 ~ ~~\forall e \in G
\label{con3}
\end{equation}
\begin{equation}
    \sum_{i|a_i \in \mathbb{A}} \phi_i^c \leq 1 ~~~\forall ~c \in C
\label{con4}
\end{equation}
\end{subequations}
%\textbf{Solution with AI:}

It is important to note that solving a constrained robust optimization problem such as \eqref{slicingopt_robust} is computationally complex. Since we use deep learning to learn the E2E QoS metrics, we may also use the same methodology to directly learn the minimized utility function over $\theta_{i}$. For example, a robust optimization strategy for throughput would be to maximize the throughput while considering the worst possible effect of $\theta_{i}$ on throughput. In general, it would be difficult to characterize the worst possible effect. However, with help of learning we can directly learn the worst possible throughput using neural networks. As discussed in Section \ref{sec:learn}, we vary the seeds for generating traffic and capture the stochastic variations in delays and throughput at the network and computing sites. For each point, we take the maximum delay and minimum throughput values for worst case modeling and feed to the neural network for training purposes. Thus, the robust optimization can be converted to a normal optimization as:
\begin{subequations}
\label{slicingopt}
\begin{equation}
    \max_{(f_i^e,\phi_i^c)} \sum_{i|a_i\in \mathbb{A}} u_i(f_i^e,\phi_i^c)
    \label{fitness}
\end{equation}
\begin{equation}
\text{s.t.} ~ D_i(f_i^e,\phi_i^c)\leq \tau_i ~~~\forall i
\label{con1}
\end{equation}
\begin{equation}
T_i(f_i^e,\phi_i^c) \geq \rho_i ~~\forall i
% T_i^N(f_i^e,\phi_i^c;\mathbf{\theta_{i}}) \geq \rho_i^N ~~~\forall ~i
\label{con2}
\end{equation}
% \begin{equation}
% T_i^S(f_i^e,\phi_i^c;\mathbf{\theta_{i}}) \geq \rho_i^S ~~~\forall ~i
% \end{equation}
% \begin{equation}
% R_i(f_i^e,\phi_i^c;\mathbf{\theta_{i}})\geq r_i ~~~\forall ~i
% \end{equation}
\begin{equation}
\sum_{i|a_i\in \mathbb{A}} f_i^e \leq 1 ~ ~~\forall e \in G
\label{con3}
\end{equation}
\begin{equation}
    \sum_{i|a_i \in \mathbb{A}} \phi_i^c \leq 1 ~~~\forall ~c \in C
\label{con4}
\end{equation}
\end{subequations}

In \eqref{slicingopt}, and foregoing discussions, we omit the stochastic parameter $\theta_{i}$ as its effect is captured using neural network. In Appendix of the paper, we show how the sequential learning and robust optimization problem may be solved in a distributed manner at cost of additional storage and learning. For optimization, any non-linear optimizer might be used. We used the sequential quadratic program as our functions were twice continuously differentiable and at each iteration, we can also obtain the lagrange multipliers required for the distributed implementation.

% \textbf{ToDo: you have not explained what you are doing here. I don't understand it, reviewers won't either. How do you train it? How do you choose the data points so that you learn the worst possible function? Please explain. Likewise, the pseudo-code below is not clear. BTW, why emphasis on SQP? Is that the only relevant or best method? Why?}

% Algorithm \ref{alg:one}, shows the algorithm used for solving the robust optmization problem \eqref{slicingopt_robust}. The utility function \eqref{eq:utility} can be devised as a function of the learnt E2E mappings which is denoted by $F_i(.)$. By learning the worst case scenarios, we directly obtain the worst possible utility as shown in Algorithm \ref{alg:one}. In this paper, we solved \eqref{slicingopt_robust} using sequential quadratic programming as the method for solving the constrained non-convex problem.

\RestyleAlgo{ruled}

\subsection{Graceful Degradation}
The problem mentioned above may be infeasible for resource-constrained scenarios. In such situations, it is desirable that the service provider be interested in relaxing the constraints such that the most prioritized applications do not violate their constraints. The notion of priority may be conceived as a mapping of the requested QoE parameters. Based on the priority, the service provider may impose appropriate penalties, thereby allowing them to gracefully degrade (relax low priority constraints to make slicing feasible) the service. We define $\psi_i(D_i()-\tau_i,T_i()-\rho_i;\theta_{i})$ as the associated penalty function for application class $a_i$. For application classes which require strict QoE satisfaction, a penalty of zero is assigned i.e., $\psi_i(.) = 0$, if $a_i$ has strict QoE requirement. Let $\mathbb{B}$ and $\mathbb{C}$ denote two sets of application classes such that those in $\mathbb{B}$ must satisfy their QoE in strict sense while those in $\mathbb{C}$ may be degraded. 

Then, the service provider may solve the following optimization to obtain a feasible set of QoE constraints that may be satisfied by the original optimization.

\begin{subequations}
\label{gracefuldegrade}
\begin{equation}
    \min_{(f_i^e,\phi_i^c)} \sum_{i|a_i\in \mathbb{C}}  \psi_i\Big(D_i()-\tau_i,T_i()-\rho_i\Big) - \sum_{i|a_i \in \mathbb{A}}u_i(f_i^e,\phi_i^c)
    \label{grafit}
\end{equation}
\begin{equation}
\text{s.t.} ~ D_i(f_i^e,\phi_i^c)\leq \tau_i ~~\forall ~i \in \mathbb{B} \label{gracon1}
\end{equation}
\begin{equation}
T_i(f_i^e,\phi_i^c) \geq \rho_i ~~\forall ~i \in \mathbb{B} \label{gracon2}
\end{equation}
\begin{equation}
\sum_{i|a_i\in \mathbb{A}} f_i^e \leq 1 ~~\forall e \in G
\label{gracon3}
\end{equation}
\begin{equation}
    \sum_{i|a_i \in \mathbb{A}} \phi_i^c \leq 1 ~~\forall ~c \in C
    \label{gracon4}
\end{equation}

% \begin{equation}
%     \tau'_i\geq\tau_i, ~ \rho'_i \leq \rho_i ~~\forall ~i
% \end{equation}
\end{subequations}

In solving this problem, the optimal slicing and feasible QoE constraints may be obtained based on the priority settings similar to the previous case.

\section{Implementation of Digital Twin}
To demonstrate a  digital twin implementation, we used an emulator named ComNetsEmu developed at Granelli Lab \cite{xiang2021open}. The emulator is developed over the traditional SDN emulator MININET where each host has docker installed for container management \cite{xiang2021open}. In order to mimic Kubernetes, ComNetsEmu uses a docker-in-docker based concept that allows managing resources within a docker host. We present emulation results with three clients. Due to the limitations of Mininet, links could not support data rates greater than 1 Gbps.
% \subsection{Emulation Results}
\label{res:emu}
\begin{figure*}
    \centering
    \subfloat[Learnt E2E delays using Neural Network on Emulation Platform]{\includegraphics[width=0.45\linewidth]{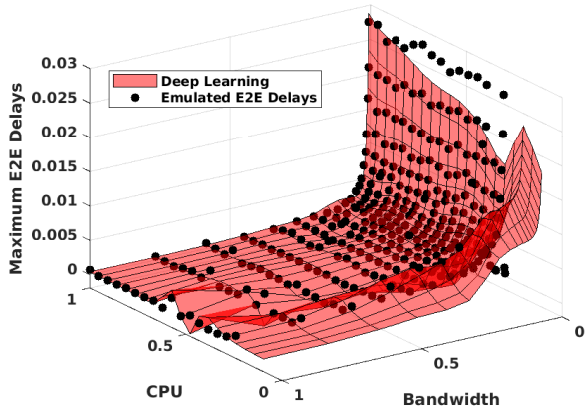}\label{fig:deep_learn}}\qquad
    \subfloat[Histogram of E2E delays of requests for Slice1 and Slice2. Delay violations show for two application.]{\includegraphics[width=0.45\linewidth]{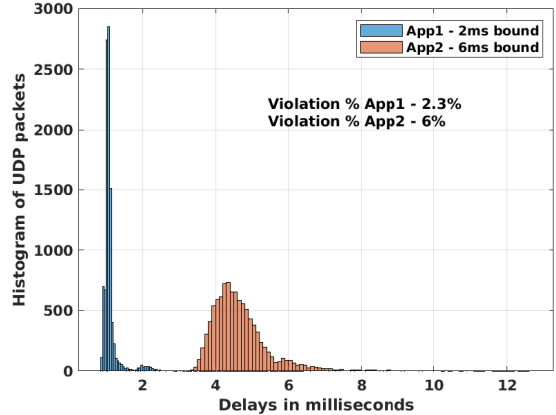}\label{fig:perfemu}}
    \caption{Figures showing implementation of sequential learning and optimization for E2E QoE satisfaction in Emulation.}
    \label{fig:my_label}
\end{figure*}
% \begin{figure}
%     \centering
%     \subfloat[Learnt E2E delays using Neural Network on Emulation Platform]{\includegraphics[width=0.45\linewidth]{Figures/deep_learn.eps}\label{fig:deep_learn}}  \\
%     \subfloat[Histogram of E2E delays of requests for Slice1 and Slice2. Delay violations show for two application.]{\includegraphics[width=0.45\linewidth]{Figures/emuvio.eps}\label{fig:perfemu}}
%     \caption{Figures showing implementation of sequential learning and optimization for E2E QoE satisfaction in Emulation.}
%     \label{fig:my_label}
% \end{figure}

We implement a radio access network setup as shown in Fig.\ref{fig:simsetup}. Three clients were connected to an edge server via an OpenFlow switch which an SDN controller controls. The propagation delays of the links are distributed uniformly between 5 $\mu$s and 25 $\mu$s (1-5 Km). Each client sends UDP packets to the server. The packets contain array data, and the server finds Fast Fourier Transforms of the data and returns the result. To differentiate between applications of different processing requirements, we consider an application (App1) which computes the transform 1000 times for a request while the other (App2) does the exact 10000 times, respectively. We set the delay constraints for the two applications to be 2ms and 6ms for the emulation setup.

The E2E delays are collected and learnt using a 2$\times$16$\times$32$\times$8$\times$1 neural network. Fig. \ref{fig:deep_learn} shows the performance of the network. The collected values are plotted over the learnt function. The figure shows that the delays exhibit non-linear nature, as mentioned before. Fig. \ref{fig:perfemu} shows the performance of joint-optimization over the learnt E2E delays. The optimal CPU and bandwidth allocations returned all sent requests, and hence the throughput for both applications was 100\%. However, as can be seen from Fig.\ref{fig:perfemu}, App1 violates delay constraints by 2.3\% while App2 does so by 6\%. The small percentage of violation is due to the combined effect of learning inefficiency as well as the non-convex nature of the optimization problem.
\section{Experimental Results}
\label{sec:expresults}
Although the emulator is quite pragmatic, large scale emulations (connecting more than ten clients) could not be supported on a laptop. To circumvent this, we created a simulator in OmNet++ based on our experience with emulation and cloud computing simulators like CloudSim \cite{calheiros2011cloudsim}. 

We conduct multiple experiments using simulation environments. These experimental scenarios are motivated by a suggestion from an actual network operator (Telstra) in the deployment of 5G networks in Australia.

% \subsection{Experiment Setup}

\subsection{Simulation Results}

\begin{table}[h]
\label{tab:params}
\centering
\caption{Parameters considered in simulations}
\renewcommand*{\arraystretch}{1.2}{
\begin{tabular}{|L{3cm}|L{3cm}|}
 \hline
\textbf{Parameter}                              & \textbf{Value} \\
\hline
(Delay, Throughput) requirements of App1 & (1ms, 90\%)                           \\
\hline

(Delay, Throughput) requirements of App2 & (5ms, 95\%)                           \\
\hline
\#Cores/Server                                                         & 2                                     \\
\hline 
Core Speed @Edge                                                       & 3e8 MIPS                              \\
\hline 
Processing required for App1                                           & 5e4 MI                                \\
\hline 
Processing required for App2                                           & 8e4 MI                                \\
\hline 
Core Speed @Cloud                                                       & 3e9 MI                               \\
\hline
\#Active Users                                                           & 10\\
\hline 
Packet generation rate                                                  & 200/second \\
\hline
Packet Arrival Process                                                  & Bursty Pareto H = 8 \\
\hline
Packet Size (Bytes)                                                            & $\sim U(20, 65535)$\\
\hline
Propagation Delay to edge (1-5 Km)                                                            & $\sim U(5 \mu s, 25\mu s) $\\
\hline
Propagation Delay from edge to cloud & 0.5 ms\\
\hline
Bandwidth between subscriber and edge (100 Km) & 1 Gbps each\\
\hline
Bandwidth between edge and cloud & 100 Gbps\\
\hline
\end{tabular}}
\end{table}

\begin{figure}
    \centering
    \includegraphics[height=8cm]{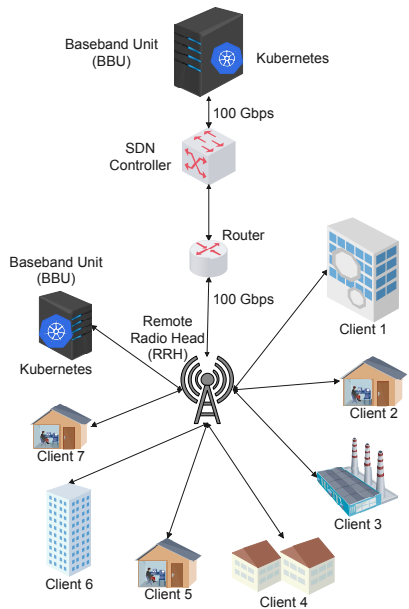}
    \caption{Simulation setting for E2E QoE constraints over RAN.}
    \label{fig:simsetup}
\end{figure}

\begin{figure*}[!t]
    \centering
    \subfloat[Performance of Deep learning approach vs Linear \cite{dietrich2017network,sattar2019optimal} and Polynomial Fit for App1]{\includegraphics[width=0.49\linewidth]{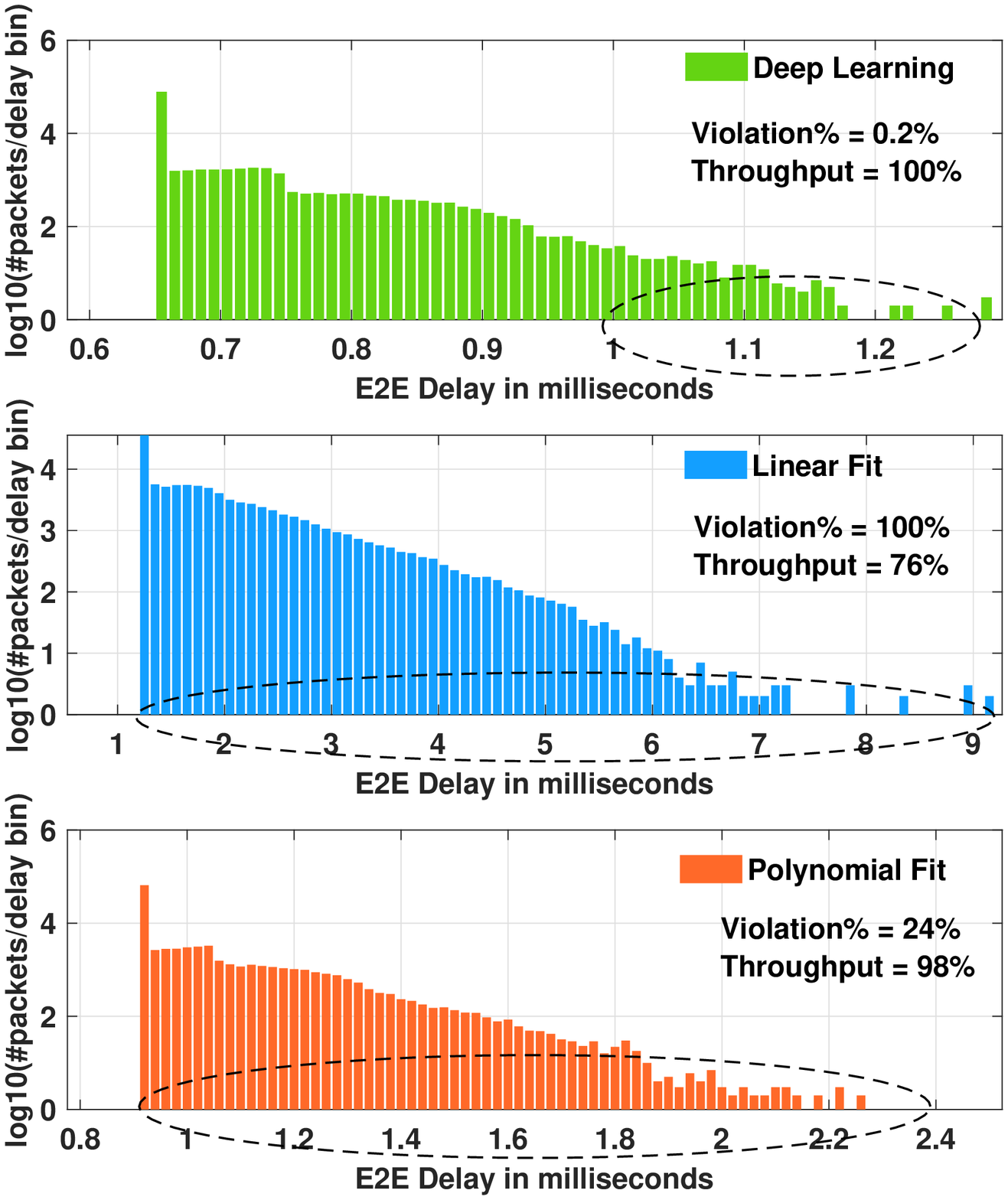}\label{fig:1ms_learn}} \hfil
    \subfloat[Performance of Deep learning approach vs Linear \cite{dietrich2017network,sattar2019optimal} and Polynomial Fit for App2]{\includegraphics[width=0.49\linewidth]{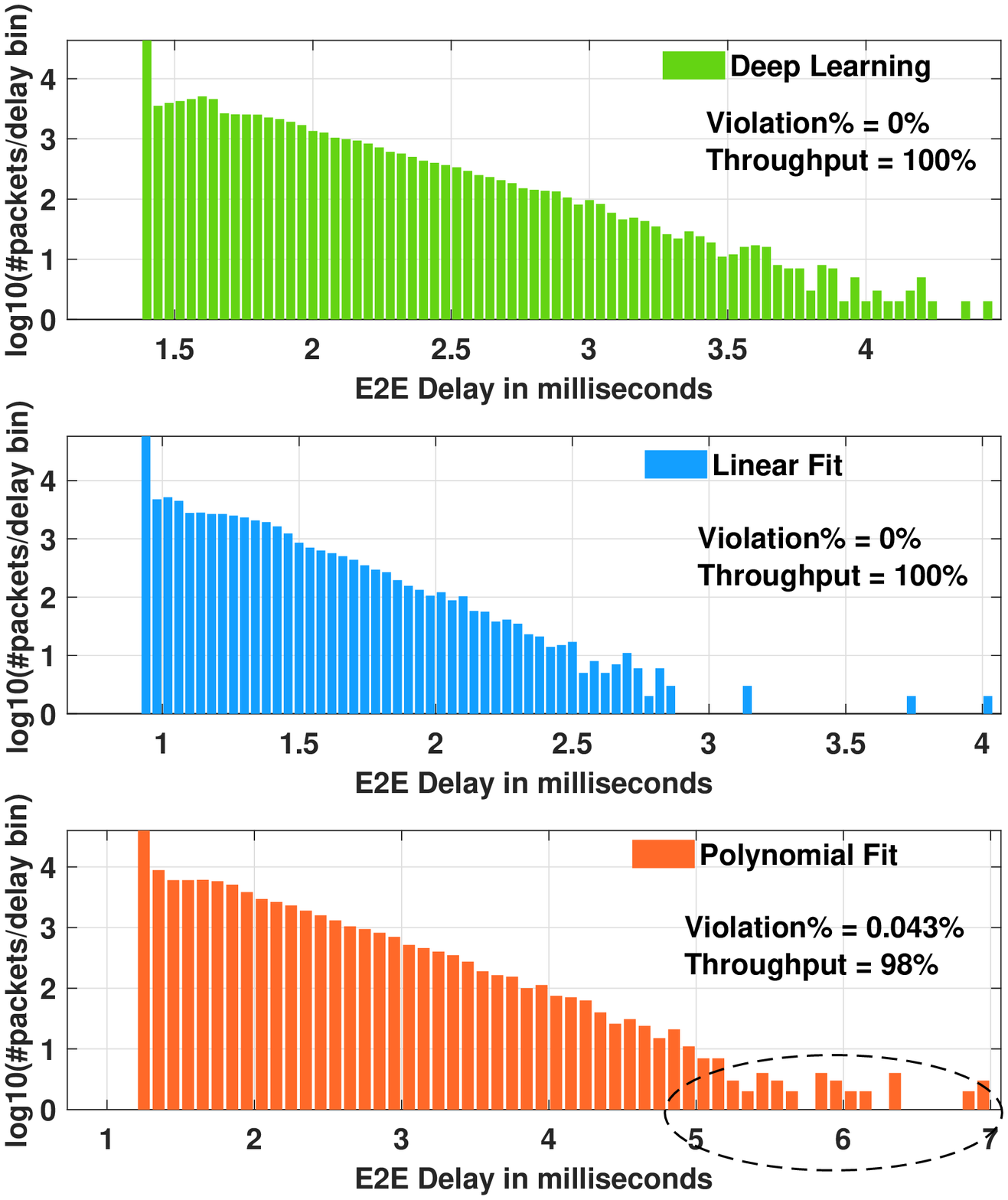}\label{fig:5ms_learn}}
    \caption{Figures showing the effectiveness of sequential deep learning over other learning approaches. Linear approach \cite{sattar2019optimal,dietrich2017network} is skewed towards one application while polynomial approach performs relatively better. Deep learning approach improves performance by a significant amount.}
    \label{fig:my_label_2}
\end{figure*}
\begin{figure*}[!t]
\centering
\subfloat[Performance of Joint Optimization vs Differentiated service for App1. ]{\includegraphics[width=0.49\linewidth]{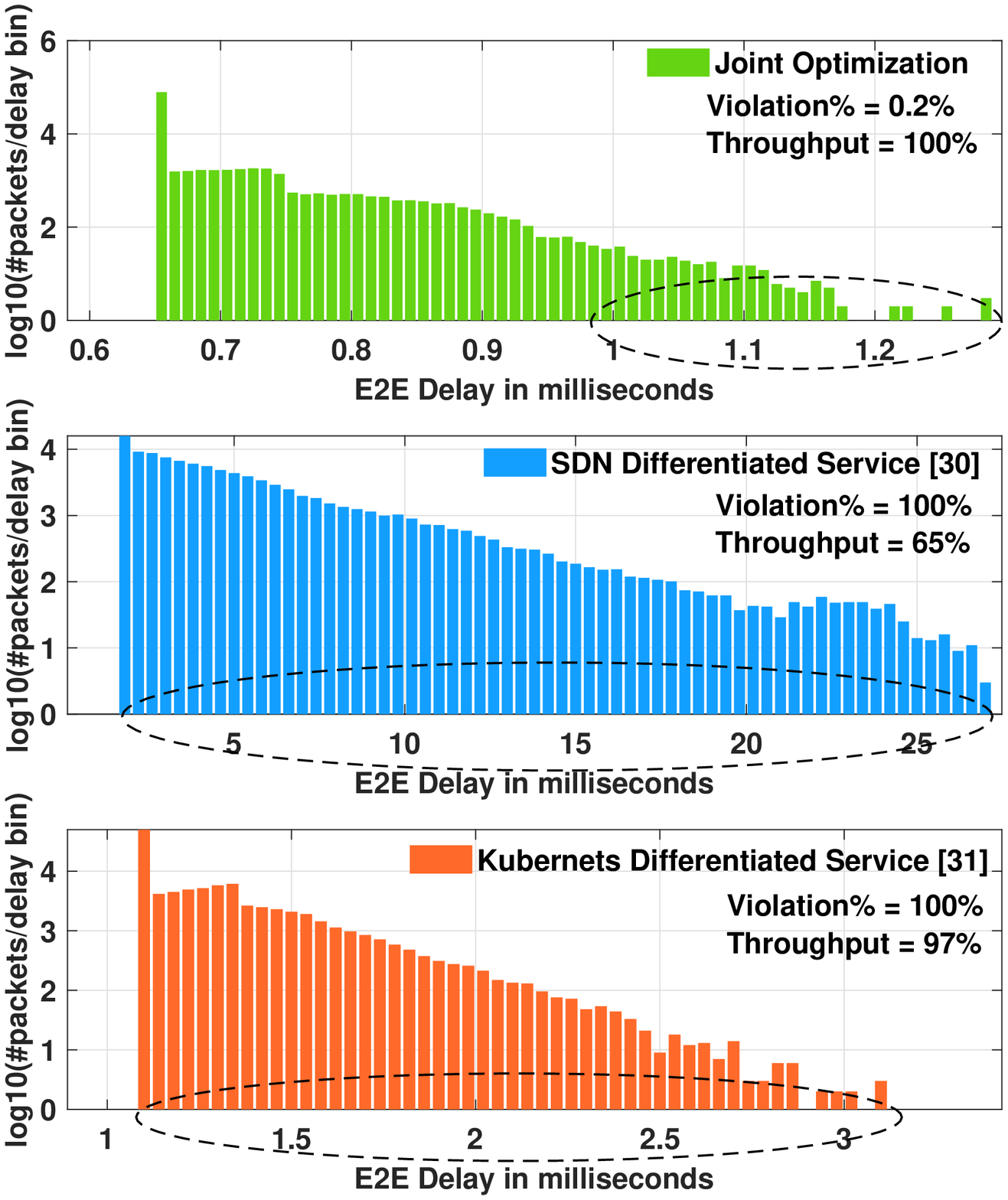}\label{fig:1ms_joint}} \hfil
\subfloat[Performance of Joint Optimization vs Differentiated service for App2. ]{\includegraphics[width=0.49\linewidth]{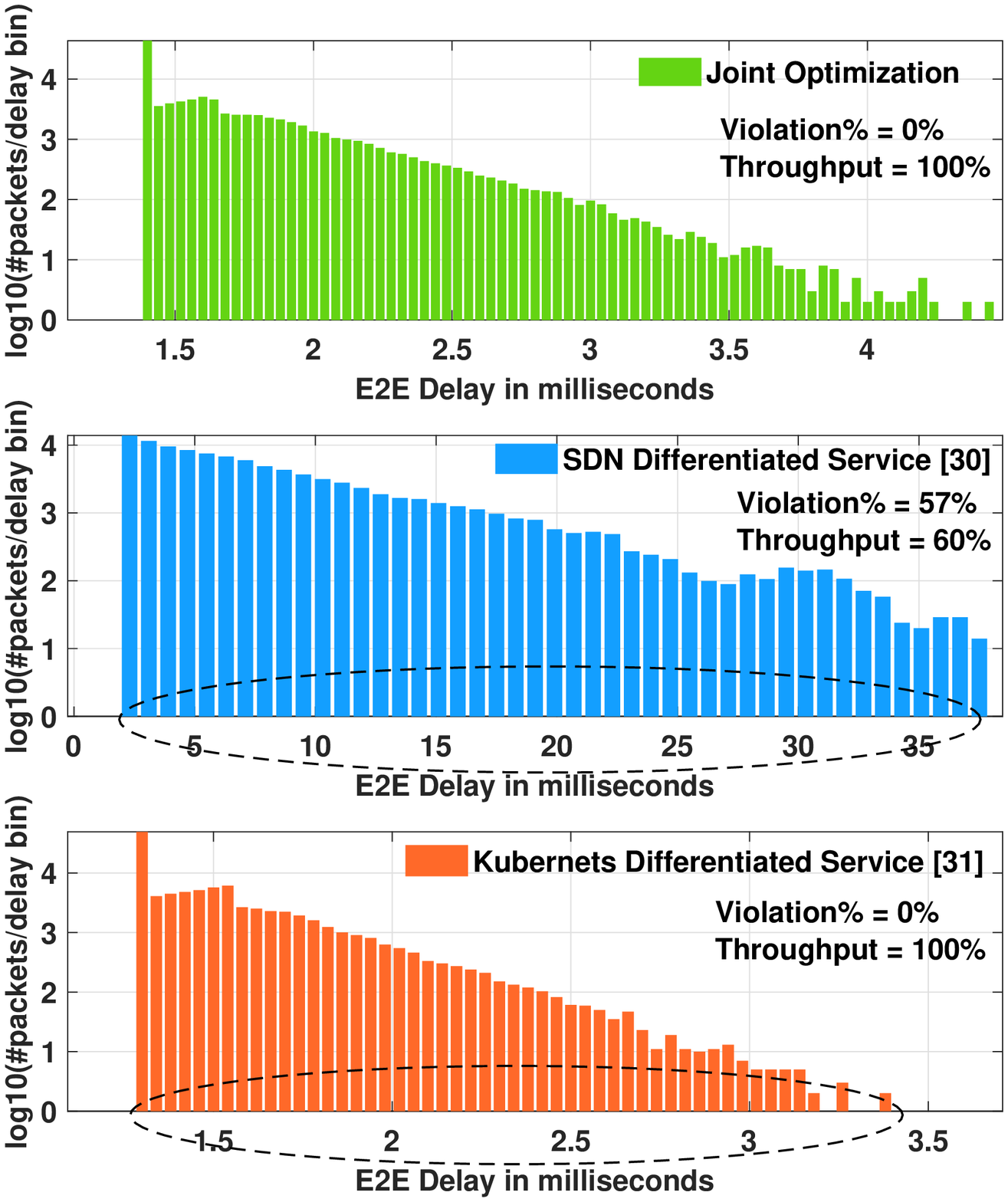}\label{fig:5ms_joint}}
\caption{Figures showing the effectiveness of sequential deep learning and joint-optimization approach over Differentiated Services. App1 is given priority at queues in the network and server sides respectively. App1 is both network and processing hungry. App2 is given priority at queues in the network and server sides respectively. App2 is more processor hungry as compared to App1 while both compete for network.}
\label{fig:mylabel3}
\end{figure*}

We present the large scale simulation studies (clients$>$10). To demonstrate the performance of our slicing approach for the physical networking scenario described in Fig. \ref{fig:simsetup}, we implement the described network in OmNet++ with ten subscribers, each requesting 200 packets per second. The traffic arrival process is bursty generated using an on-off Pareto distribution. We have two types of applications which are referred to as App1 and App2, respectively. For App1, each packet sends a request to the server to execute an operation with an average of $5\times 10^4$ Million instructions (MI) to be executed at an edge server that operates at a speed of $3 \times 10^8$ million instructions per second (MIPS). Similarly, for App2, we have a higher computing requirement of $8 \times 10^4$ MI. The end to end delay constraints for the two applications are set at 1ms and 5ms, respectively, while the throughput requirements are set at 95\% and 97\%, respectively. It is easily understood that the slice of App1 cannot be implemented at the Cloud since the round trip delay is already 1ms, and no delay $<1$ ms can be achieved at Cloud.  In order to satisfy these requirements, we create a network slice out of the total bandwidth $100$ Gbps and total computing at the edge server. The bandwidth slice is implemented by the SDN, while Kubernetes implements the slice at the edge server. 

% We present the results in a normalized scale where is normalization concerns the maximum capacities.

\subsection{Efficacy of Deep Learning}
To evaluate the performance of our slicing method, we measure the end to end delays and throughput on receiving the output at the client-side. By throughput, we mean the percentage of requests that was returned to the client successfully. We plot the histogram of packets with respect to end to end delays experienced by them. It is essential to observe that App2 resembles a heavy computation application compared to App1, which has moderate computing and networking requirements.

\begin{figure}[!t]
    \centering
    \includegraphics[height=8cm,width=8cm]{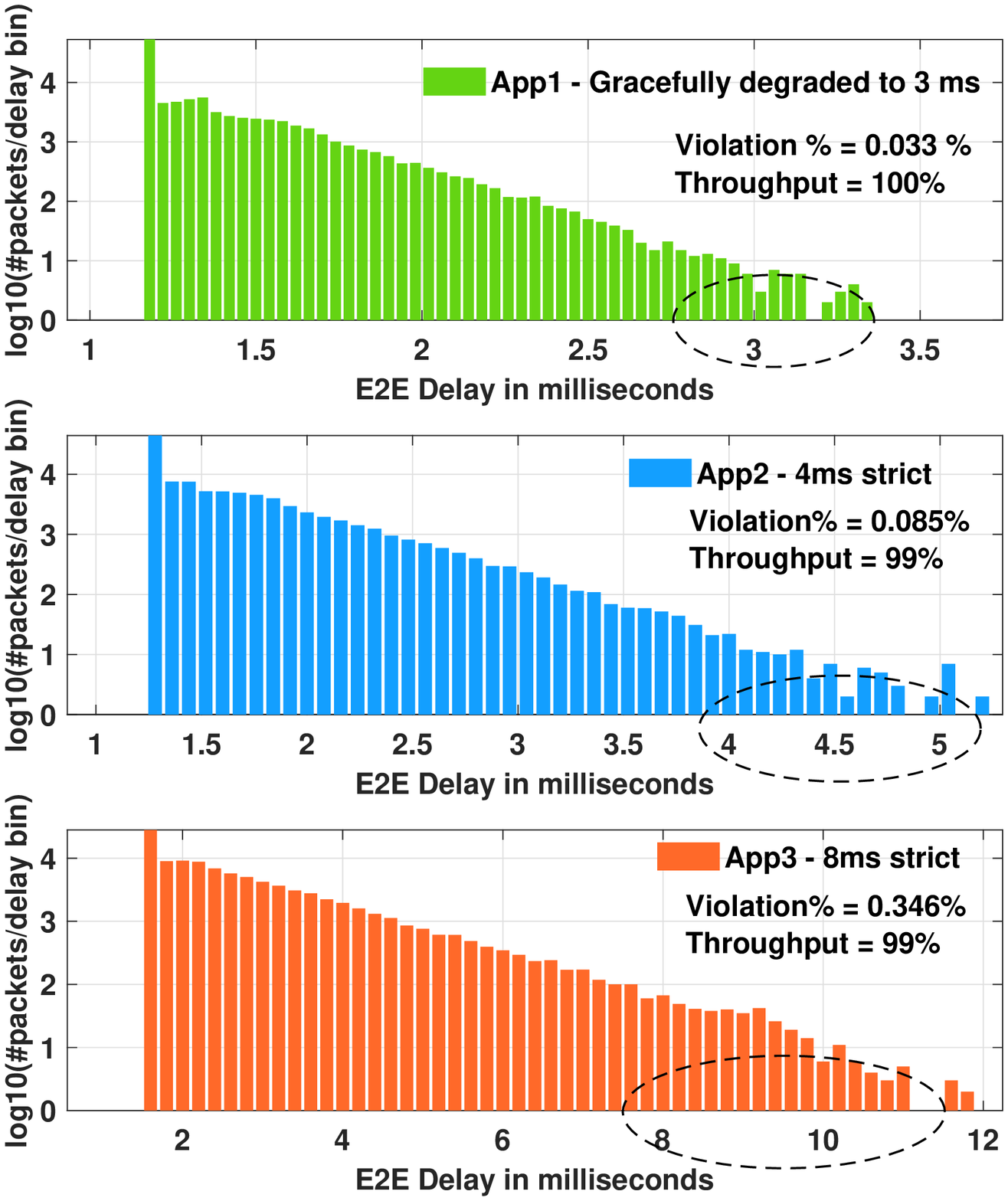}
    \caption{App3 is introduced. App2 and App3 have strict delay constraints. App1 is gracefully degraded to 3 ms.}
    \label{fig:grace}
\end{figure}

To compare our method, we consider the works of \cite{dietrich2017network,sattar2019optimal} where a linear relationship was used to estimate the delays. However, it is essential to note that these works addressed the problem of placement of applications at servers and did not deal with the processor allocation while considering the modern deployment strategies in Dockers and Kubernetes. We also compare the obtained slicing decisions with a polynomial fit instead of a deep neural network used for our approach. The polynomial used is of degree 5 in one dimension and degree 4 in the other. 

The linear relationship proposed in the literature does not yield a feasible solution for App1 while we are able to obtain feasible solutions for App2 (satisfies the constraints) whose performances can be observed from the throughput percentages and end to end delay histograms. As seen from Fig. \ref{fig:1ms_learn} and Fig. \ref{fig:5ms_learn}, our approach shows that the number of packets violating the end to end delay constraints are significantly less compared to polynomial estimates. This is presumably due to the efficacy of deep neural network as an universal function approximator. We observe that in our approach none of the packets requesting App2 violate the delay constraints while some of those of App1 do violate. This may be due to prediction errors and is marginal as shown in Fig. \ref{fig:1ms_learn}. Deep learning approach promises to achieve uRLLC requirements of satisfying strict delay constraints for 99.99\% of time. While this is exactly the case for App2, for App1 the same happens for 99.2\% of times. Linear Fit approach goes towards either extreme (infeasible allocations in one and allocates all resources to other) due to inefficient approximation of the E2E QoE functions. 

\subsection{Joint Optimization vs Differentiated Service}
To demonstrate the effectiveness of the joint-optimization technique proposed in the paper, we compare the same with prioritization based service, which is the standard way of dealing with delay constrained services \cite{cisco_2017,kubernetes_2021}. In our simulations, we preempt the prioritized App over other Apps requests either at the networking site or server site. Fig. \ref{fig:1ms_joint} shows the E2E delay performance of received packets for $10^5$ packets of App1 (1ms bound). As can be seen from Fig. \ref{fig:1ms_joint}, the number of packets that violate the delay constraint is only 0.2\% as compared to the network prioritized and computing prioritized approaches. This is due to the nature of the application. The result also suggests that the App1 is sensitive to both network and computing. Only the joint-optimization technique can effectively handle its QoE requirements. Hence in either differentiated methods, we observe that the obtained delays violate the required bounds. 

Fig. \ref{fig:5ms_joint} shows the E2E delay performance of received packets for $10^5$ packets of App2 (5ms bound). As can be seen from Fig. \ref{fig:1ms_joint}, none of packets violate the delay constraint compared to the network prioritized approach. This is again due to the nature of the application. The result suggests that the App2 is  sensitive to computing. In this case, assigning more computing power helps in achieving the required delay constraint.

\subsection{Graceful Degradation}
We introduce a new App - App3 with 8e4 MI and a delay constraint of $8ms$, which returns an infeasible solution (constraint satisfaction is not possible). In such cases, our approach proposes using the graceful degradation framework, which aims to find minimum possible relaxations on the requirements. In this regard, we want App2 and App3 to strictly satisfy its constraints while imposing a penalty on the deviation of the App1 delay constraint in our objective. As we observe from Fig. \ref{fig:grace}, the optimization finds the feasible delay constraint to be 3ms. However, we observe that the new slicing decisions have increased the number of packets violating delay for App3. This might be due to involved errors in learning the functions.

% \textcolor{red}{Fig \ref{fig:grace} not cited in text.}

\section{Conclusion}
\label{sec:sconclusion}
In this paper, we use deep learning to learn the relationship between resources and E2E QoE metrics for slicing in radio access networks (RAN). Deep learning proves to be highly efficient in modelling non-linear relationships compared to polynomial fitting approaches and linear approximations used in literature. In addition, learning helps in simplifying a robust optimization problem to a conventional optimization one. We also show how the learning and optimization can be implemented in a distributed manner on network and server controllers at cost of additional memory. Further, we show how we can make resource allocation decisions in resource-constrained scenarios by gracefully degrading a slice. However, learning approach is not free from errors and its difficult to ensure bounds rather than practical implementations. In this regard, conservative design approach may be employed i.e. to find allocations for stricter bounds than what is agreed upon. In this work, we only considered E2E delays and throughput as QoE metrics. We can also consider other relevant QoE metrics, such as the jitter. Large scale studies over servers and networks with such frameworks can be carried out. Theoretical extensions could involve finding bounds on the robustness of thus found optimal solutions.

\section*{Acknowledgment}
The authors would like to thank Mr. Akilan Wick from Telstra Australia for his useful inputs regarding the Australian 5G deployment scenario.
\appendices
\section{Distributed Networking and Computing Implementation}
As discussed in Section \ref{sec:introduction}, for service providers without the capability of implementing a digital twin, it is desirable that the SDN and Kubernetes achieve E2E QoE in a distributed manner. We observe that the decision variables of the two entities are different. The SDN has the responsibility to decide $f_i^e$ while Kubernetes decides $\phi_i^c$. We turn our attention to  constraints. 

% \subsubsection{Separable Objectives}
% It is understood that SDN can put its own objective (say $U^N(f_i^e)$) and Kubernetes can put its own (say $U^K(\phi_i^c)$). Also, the priority based penalty function can be defined by SDN and Kubernetes separately and added to obtain the overall penalties.
% \subsubsection{Separable Constraints}
% Network Throughput and Computing Throughput are already separate constraints in their individual variables. Same is the case for the link capacity and server capacity constraints.
% \subsubsection{Complicating Constraints}
It is evident that the E2E delay and throughput functions depend on both variables. Due to additive nature of delays, the E2E delay can be expressed as sum of delay functions of network and computing site respectively. Note that E2E throughput is determined by number of successful requests. A request is successful if it is successfully transferred by the network and subsequently processed at the server. The two events of success at network and at server are independent and hence their probabilities can be multiplied. To convert this into a sum, we may use a logarithm operation over the throughput function. Further, the randomness associated with the random variable $\theta_{i}$ can be accounted using equations \eqref{eq:delsum},\eqref{eq:thruprod}.   
\begin{align}
    \forall \theta_{i},~ D_i(f_i^e,\phi_i^c;\theta_{i}) & =   D_i^N(f_i^e;\theta_{i})+ D_i^S(\phi_i^c;\theta_{i}) \label{eq:delsum}\\
    \forall \theta_{i},~T_i(f_i^e,\phi_i^c;\theta_{i}) & =  T_i^N(f_i^e;\theta_{i}) \times T_i^S(\phi_i^c;\theta_{i})
    \label{eq:thruprod}
\end{align}
% \subsection{Decomposition Technique}
\begin{figure*}[!t]
    \centering
    \subfloat[Learning at SDN and Kubernetes with separate neural networks]{
    \includegraphics[height=10cm,angle=90]{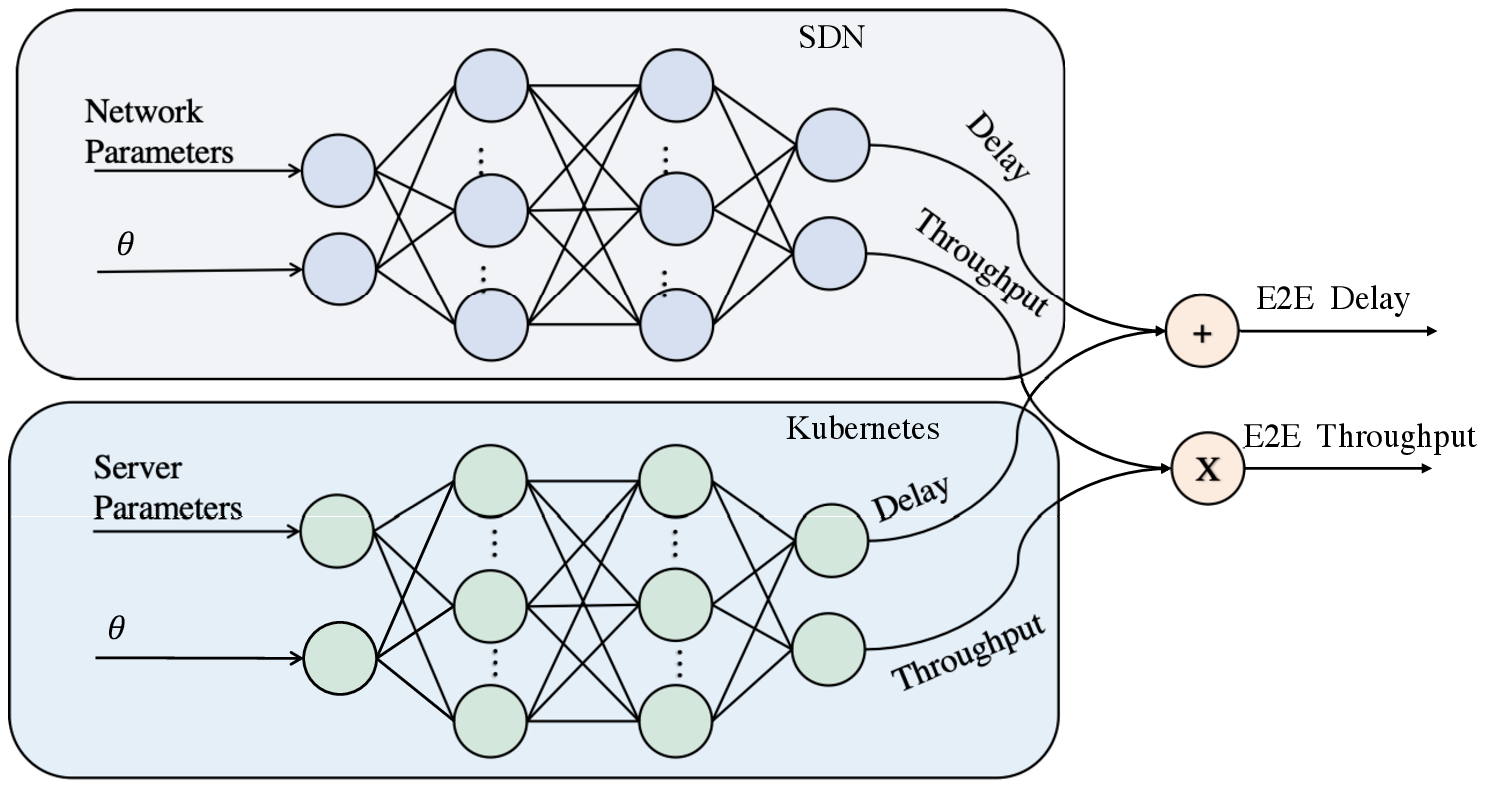}
    \label{fig:distlearn}}\hfil
    \subfloat[Overview of Distributed learning and optimization technique.]{\includegraphics[height=8cm]{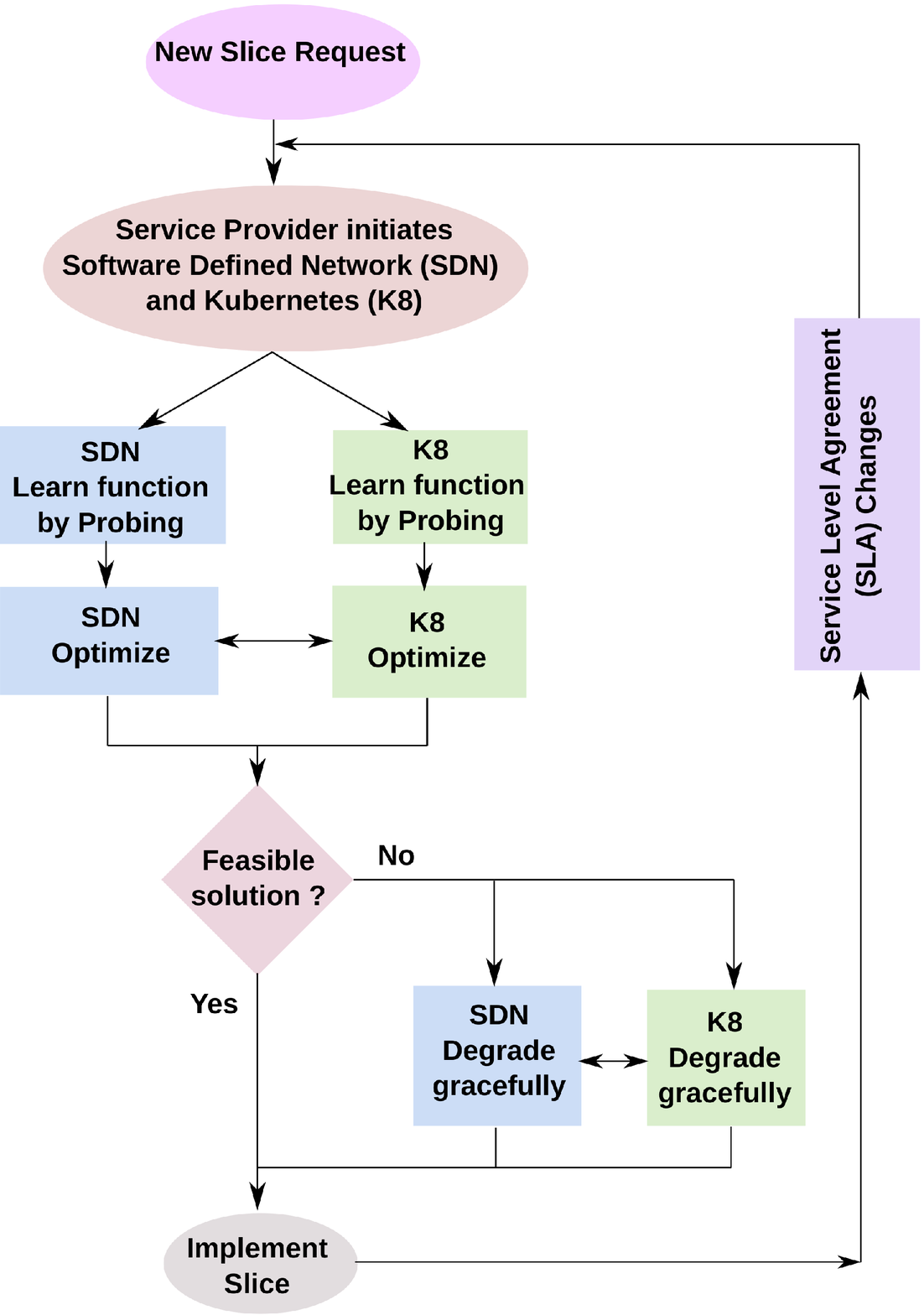}
    \label{fig:distalgo}}
    \caption{Distributed Implementation of Sequential Learning and Robust Optimization in SDN and Kubernetes}
\end{figure*}
As shown in Fig. \ref{fig:distlearn}, the SDN obtains the network delays experienced by the packets when packets flow within the network via the routers. One can implement a network monitoring application \cite{Software84:online} to collect relevant delay and drop statistics. Using a network hypervisor like Flowvisor, the network can be sliced with different bandwidth configurations \cite{xiang2021open} for the packets corresponding to a specific application. Similar to the SDN, Kubernetes \cite{Managing53:online} can vary the computing power allocated to a container (application) and calculate the delays and drops (crashes/out of memory) to execute the container, if any. To account for the stochastic nature of network state and server, service provider uses a Monte-Carlo strategy (by using different random seeds to generate traffic) and obtain the required statistics. For example, for a given bandwidth and processing configuration, one can obtain the histogram for E2E delays experienced by the packets. Once the data regarding the QoE metrics are collected by the SDN and Kubernetes, they may train individual neural networks as shown in Fig. \ref{fig:distlearn}. The  E2E delay and throughput may be computed using \eqref{eq:delsum} and \eqref{eq:thruprod}.
% \begin{figure}[h]
%     \centering
%     \includegraphics[height=10cm,width=10cm]{Figures/flow-slicing_v2.eps}
%     \caption{Proposed approach for network slicing with learning and optimization phases.}
%     \label{fig:distalgo}
% \end{figure}
Although in case of centralized robust solution, we could learn the worst case delays directly, the same cannot be done for a distributed approach. This is due to the fact that maximum or minimum are non-linear operators. Thus, for each realization of $\theta_{i}$, the delays and throughput values are computed by SDN and Kubernetes and fed into the neural networks at corresponding sites. We use primal decomposition technique for distributed optimization \cite{boyd2007notes}. The overall algorithm is shown in Fig. \ref{fig:distalgo}.
% As discussed in \ref{sysmod}, it is desirable that the SDN and Kubernetes solve the slicing problem in a distributed manner. To facilitate, we first observe that the decision variables of the two entities are different. The SDN has the responsibility to decide $f_i^e$ while Kubernetes decides $\phi_i^c$. We turn our attention to the objectives and constraints. 
% % \subsubsection{Separable Objectives}
The decision variables of the two entities - SDN and Kubernetes are different, namely $f_i^e$ and $\phi_i^c$ respectively. SDN may have its own objective (say $u_i^N(f_i^e)$) and Kubernetes may have another one (say $u_i^S(\phi_i^c)$). These objectives may or may not depend on $\theta_{i}$. In this section, we present the simpler case where these are independent of $\theta_{i}$ and present the more general version in the next subsection of the paper. 
\subsection{Distributed Robust Optimization without Stochastic Objective}
As presented in \eqref{eq:delsum} and \eqref{eq:thruprod}, the E2E delay and E2E throughput are separable as for a given $\theta_{i}$. For decentralized implementation, it is important that we decompose the functions as sums instead of a product. This can be easily achieved for E2E throughput by taking a logarithm on both sides of (\ref{eq:thruprod}): 

% Furthermore, the priority based penalty function can be defined by SDN and Kubernetes separately and added to obtain the overall penalties. 
 
% % \subsubsection{Separable Constraints}
% % Network Throughput and Computing Throughput are already separate constraints in their individual variables. Same is the case for the link capacity and server capacity constraints.
% % \subsubsection{Complicating Constraints}
% It is evident that the E2E delay and throughput functions depend on both variables. Due to additive nature of delays the E2E delay can be expressed as sum of delay functions of network and computing site respectively. Note that E2E throughput is determined by number of successful requests. A request is successful if it is successfully transferred by the network and subsequently processed at the server. The two events of success at network and at server at independent and hence their probabilities can be multiplied. To convert this into a sum, we use a logarithm operation over the throughput function.   
% \begin{equation}
%     D_i(f_i^e,\phi_i^c) = D_i^N(f_i^e)+D_i^S(\phi_i^c)
% \end{equation}
\begin{equation}
    \forall \theta_{i}, ~ \log T_i(f_i^e,\phi_i^c,\theta_{i}) = \log T_i^N(f_i^e,\theta_{i})+ \log T_i^S(\phi_i^c,\theta_{i})
\end{equation}
% % \subsection{Decomposition Technique}
% For ease of notational convenience, we omit the $\log$ operation in the following equations. 

We introduce variables $\tau_i^N$ and $\rho_i^N$ such that:

$$D_i^N(f_i^e,\theta_{i})\leq \tau_i^N \text{ and } D_i^S(\phi_i^c,\theta_{i})\leq \tau_i-\tau_i^N$$
$$\log T_i^N(f_i^e,\theta_{i})\leq \log \rho_i^N \text{ and } \log T_i^S(\phi_i^c,\theta_{i}) \leq \log \rho_i -\log \rho_i^N$$
Thus for a given $\theta_{i},\tau_i^N,\rho_i^N$, the SDN and Kubernetes can solve their own optimization problems as:\\

\noindent\begin{minipage}{.5\linewidth}
\textbf{P(N)[SDN]:}

\begin{subequations}
\label{sdnopt}
\begin{equation*}
    \max_{f_i^e} \sum_{i|a_i\in \mathbb{A}} u_i^N(f_i^e)
\end{equation*}
\begin{equation*}
   \text{s.t. } ~ D_i^N(f_i^e,\theta_{i})\leq \tau_i^N
\end{equation*}
\begin{equation*}
    \log T_i^N(f_i^e,\theta_{i})\leq \log \rho_i^N
\end{equation*}
% \begin{equation}
%     R_i^N(f_i^e) \geq r_i^N
% \end{equation}
\begin{equation*}
\sum_{i|a_i\in \mathbb{A}(e)} f_i^e \leq 1 ~ ~~\forall e \in G
\end{equation*}
\end{subequations}
\end{minipage}%
\begin{minipage}{.5\linewidth}
\textbf{P(S)[Kubernetes]:}

\begin{subequations}
\label{kuberopt}
\begin{equation*}
    \max_{\phi_i^c} \sum_{i|a_i\in \mathbb{A}} u_i^S(\phi_i)
\end{equation*}
\begin{equation*}
   \text{s.t. } ~ D_i^S(\phi_i^c,\theta_{i})\leq \tau_i-\tau_i^N
\end{equation*}
\begin{equation*}
    \log T_i^S(\phi_i,\theta_{i})\leq \log \rho_i - \log \rho_i^N 
\end{equation*}
% \begin{equation}
%     R_i^S(\phi_i^c) \geq r_i-r_i^S
% \end{equation}
\begin{equation*}
    \sum_{i|a_i \in \mathbb{A}} \phi_i^c \leq 1 ~~~\forall ~c \in C
\end{equation*}
\end{subequations}
\end{minipage}\\

On solving these, the SDN and Kubernetes obtains the optimal $f_i^e,\phi_i^c$ for a given $(\tau_i^N,\rho_i^N,\theta_{i})$. We may express the optimal value of the utilities by $N(\tau_i^N,\rho_i^N,\theta_{i})$ and $S(\tau_i^N,\rho_i^N,\theta_{i})$. Finally the objective is the maximize the sum of the obtained functions which is often termed as the master problem\cite{boyd2007notes}:
\begin{equation}
    \max_{(\tau_i^N,\rho_i^N,\theta_{i})} N(\tau_i^N,\rho_i^N,\theta_{i}) + S(\tau_i^N,\rho_i^N,\theta_{i})
    \label{eq:master}
\end{equation}
Note that $\tau_i^N \in[0,\tau_i]$, $\rho_i^N \in[0,\rho_i]$ and $\theta_{i} \in \theta_{i}$. Let us denote the domains for $\tau_i$ and $\rho_i$ by $\mathbb{T}$ and $\mathbb{P}$. Let  $\Pi_{\mathbb{C}}(\bf{x})$ denote the projection of the point $\bf{x}$ over set $\mathbb{C}$. The sub-gradient update equations for $(\tau_i^N,\rho_i^N,\theta_{i})$ are:
\begin{subequations}
\label{eq:grad_update}
\begin{equation}
    \tau_i^N:=\Pi_\mathbb{T}\Big(\tau_i^N+\alpha(\lambda_\tau^N-\lambda_\tau^S)\Big)
\end{equation}
\begin{equation}
    \rho_i^N:=\Pi_\mathbb{P}\Big(\rho_i^N+\alpha(\lambda_\rho^S-\lambda_\rho^N)\Big)
\end{equation}
\begin{equation}
    \theta_{i}:=\Pi_{\theta_{i}}\Big(\theta_{i}+\alpha(g_{\theta_{i}}^N+g_{\theta_{i}}^S)\Big)
\end{equation}
\end{subequations}
$\lambda_\tau^N,\lambda_\tau^S,\lambda_\rho^S,\lambda_\rho^N$ denote the Lagrange multipliers associated with the constraints w.r.t to $\tau$, $\rho$ at SDN and Kubernetes respectively. $g_{\theta_{i}}^N$, $g_{\theta_{i}}^S$ denotes the sub-gradients w.r.t $\theta_{i}$ of the formed Lagrangians at SDN and Kubernetes respectively. Although, we wrote the generalized projection operators, the operators are quite straightforward for the sets described above. For, $\tau_i^N\in[0,\tau_i]$ the projection operator is simply the minimum of the obtained value and $\tau_i$ or else zero in case the obtained value is negative. The same is the case for $\rho_i^N$. For $\theta_{i}$ which models the arrival process, the obtained value must be non-negative at each step. 
% Although we only wrote the objectives in terms of utility functions, the same idea can be extended to the penalty based graceful degradation as well. 

Algorithm \ref{alg:dist} shows the steps for distributed optimization. As starting points, we choose $\tau_i^N = \frac{\tau_i}{2}$, $\rho_i^N = \frac{\rho_i}{2}$, $\theta_{i} = \bar\theta_{i}$. $\bar\theta_{i}$ denotes the average arrival rate. We have well defined problems at SDN and Kubernetes sites which are solved by them in parallel. We use sequential quadratic programming which also provides us with the Lagrange multipliers. These are to be exchanged among SDN and Kubernetes after each iteration. The algorithm stops if the gradients (difference of lagrange multipliers) are less than some predefined small number $\epsilon$. It is to be noted that the obtained solution is locally optimal in case of generalized functions and global optimal is achieved under convexity assumptions.
% In case of robust optimization, it is understood that the parameter $\theta_{i}$ will be a shared one which needs to be communicated among the two along with some variables during an iterative optimization process.
\begin{algorithm}[hbt!]
\caption{Algorithm for Distributed Implementation}\label{alg:dist}
\KwIn{$\epsilon$, $\tau_i^N = \frac{\tau_i}{2}$, $\rho_i^N = \frac{\rho_i}{2}$,$\theta_{i} = \bar\theta_{i}$ Objectives and Constraints of P(N) and P(S)}
% \KwIn{Known function to define Penalities and Utility in terms of E2E QoE -  $\psi_i(.),F_i(.) ~~\forall i$  \newline 
% QoE Mappings from Learning - $D_i(f_i^e,\phi_i^c;\mathbf{\theta_{i}}),T_i(f_i^e,\phi_i^c;\mathbf{\theta_{i}}),\newline
% \max\limits_{\theta_{i}} D_i(f_i^e,\phi_i^c;\mathbf{\theta_{i}}),\min\limits_{\theta_{i}}T_i(f_i^e,\phi_i^c;\mathbf{\theta_{i}})$}
\KwOut{$f_{i}^{*e},\phi_{i}^{*c} ~~\forall ~i$}
% \textbf{Compute:}\newline 
% $\max\limits_{\theta_{i} \in \theta_{i}} \psi_i(.;\theta_{i}) -  u_i(.;\theta_{i}) = \psi_i\Big(\max\limits_{\theta_{i}} D_i(f_i^e,\phi_i^c;\mathbf{\theta_{i}}) - \tau'_i,\min\limits_{\theta_{i}}T_i(f_i^e,\phi_i^c) - \rho'_i)\Big) - F\Big(\max\limits_{\theta_{i}} D_i(f_i^e,\phi_i^c;\mathbf{\theta_{i}}),\min\limits_{\theta_{i}}T_i(f_i^e,\phi_i^c)\Big), ~~\forall ~i$ \;
% \textbf{Solve \ref{gracefuldegrade} using SQP:}\newline
% $f_{i}^{*e},\phi_{i}^{*c},q'^*_i = \argmax \ref{grafit} \text{ s.t.} \ref{gracon1},\ref{gracon2},\ref{gracon3},\ref{gracon4}$\;
% $X \gets x$\;
% $N \gets n$\;
\While{True}{
\textit{\textbf{@SDN}:
SDN solves P(N) to obtain $f_i^e$ and optimal Lagrange multipliers $\lambda_\tau^N,\lambda_\rho^N,g_{\theta_{i}}^N$\;
\textbf{@Kubernetes:}
Kubernetes solves P(S) to obtain $\phi_i^c$ and optimal lagrange multipliers} $\lambda_\tau^S,\lambda_\rho^S,g_{\theta_{i}}^S$\;
Exchange Multipliers and sub-gradients\;
Update for Master Problem using \eqref{eq:grad_update}\;
Stopping Criteria: $|\lambda_\tau^N-\lambda_\tau^S|$ \& $|\lambda_\rho^N-\lambda_\rho^S|$ \& $(g_{\theta_{i}}^N+g_{\theta_{i}}^S) <\epsilon$\;

% \textbf{Exchange}  $\lambda_\tau^N,\lambda_\rho^N$\ and $\lambda_\tau^S,\lambda_\rho^S$\;
% \textbf{Update 1 if $|\lambda_\tau^N-\lambda_\tau^S|>\epsilon$:} 
% $\tau_i^N:=\tau_i^N+\alpha_k^1(\lambda_\tau^N-\lambda_\tau^S)$\;
% \textbf{Update 2 if $|\lambda_\rho^S-\lambda_\rho^N|>\epsilon$:} $\rho_i^N:=\rho_i^N+\alpha_k^2(\lambda_\rho^S-\lambda_\rho^N)$\;

% % \If{$|\lambda_\tau^N-\lambda_\tau^S|<\epsilon$}
% %   {
% %     Skip Update1\;
% %   }
% %   \If{$|\lambda_\rho^S-\lambda_\rho^N|<\epsilon$}
% %   {
% %       Skip Update2\;
% %     }
% \textbf{Stopping Criteria:}\newline
%   \If{$|\lambda_\tau^N-\lambda_\tau^S|<\epsilon ~\&~ |\lambda_\rho^S-\lambda_\rho^N|<\epsilon$ }
%   {
%     break\;
%   }
  
}
\end{algorithm}
\subsection{Distributed Robust Optimization with Stochastic Objective}
In this section, we discuss the solution to the generalized distributed robust optimization problem where the objective function depends on the stochastic parameter $\theta_{i}$. The optimization problem to be solved in distributed manner is that of \eqref{slicingopt_robust}, where the objective function contains the stochastic parameter $\theta_{i}$. For example, utility is throughput, as considered in the paper. However, it is essential that SDN and Kubernetes design their own objectives which are coupled by the variable $\theta_{i}$ i.e.
\begin{equation}
    u_i(f_i^e,\phi_i^c,\theta_{i}) = u_i^N(f_i^e,\theta_{i}) + u_i^S(\phi_i^c,\theta_{i}) 
\end{equation}
For a given, $\theta_{i}$, the problem is again separable, with separate objectives for SDN and Kubernetes. Since the aim is to perform robust optimization i.e. maximizing the utility over the worst possible effect of $\theta_{i}$ i.e. minimized over $\theta_{i}$. However, on fixing $\theta_{i}$, the only optimization variables left are $f_i^e$ and $\phi_i^c$ and hence the sub-problems solved by SDN and Kubernetes are still the same.\\

\noindent\begin{minipage}{.5\linewidth}
\textbf{P(N)[SDN]:}

\begin{subequations}
\label{sdnopt}
\begin{equation*}
    \max_{f_i^e} \sum_{i|a_i\in \mathbb{A}} u_i^N(f_i^e,\theta_{i})
\end{equation*}
\begin{equation*}
   \text{s.t. } ~ D_i^N(f_i^e,\theta_{i})\leq \tau_i^N
\end{equation*}
\begin{equation*}
    \log T_i^N(f_i^e,\theta_{i})\leq \log \rho_i^N
\end{equation*}
% \begin{equation}
%     R_i^N(f_i^e) \geq r_i^N
% \end{equation}
\begin{equation*}
\sum_{i|a_i\in \mathbb{A}(e)} f_i^e \leq 1 ~ ~~\forall e \in G
\end{equation*}
\end{subequations}
\end{minipage}%
\begin{minipage}{.5\linewidth}
\textbf{P(S)[Kubernetes]:}

\begin{subequations}
\label{kuberopt}
\begin{equation*}
    \max_{\phi_i^c} \sum_{i|a_i\in \mathbb{A}} u_i^S(\phi_i,\theta_{i})
\end{equation*}
\begin{equation*}
   \text{s.t. } ~ D_i^S(\phi_i^c,\theta_{i})\leq \tau_i-\tau_i^N
\end{equation*}
\begin{equation*}
    \log T_i^S(\phi_i,\theta_{i})\leq \log \rho_i - \log \rho_i^N 
\end{equation*}
% \begin{equation}
%     R_i^S(\phi_i^c) \geq r_i-r_i^S
% \end{equation}
\begin{equation*}
    \sum_{i|a_i \in \mathbb{A}} \phi_i^c \leq 1 ~~~\forall ~c \in C
\end{equation*}
\end{subequations}
\end{minipage}\\

On solving these, the SDN and Kubernetes obtains the optimal $f_i^e,\phi_i^c$ for a given $(\tau_i^N,\rho_i^N,\theta_{i})$. We may express the optimal value of the utilities by $N(\tau_i^N,\rho_i^N,\theta_{i})$ and $S(\tau_i^N,\rho_i^N,\theta_{i})$ as before. Hoever, the master problem now changes to:
\begin{equation}
    \max_{(\tau_i^N,\rho_i^N)\in \mathbb{T}\times\mathbb{P}} \min_{\theta_{i} \in \theta_{i}} ~~ N(\tau_i^N,\rho_i^N,\theta_{i}) + S(\tau_i^N,\rho_i^N,\theta_{i})
    \label{eq:master}
\end{equation}

This problem is equivalent to finding saddle-points or commonly known as the saddle-point problem\cite{nedic2009subgradient}.
The sub-gradient update equations for $(\tau_i^N,\rho_i^N,\theta_{i})$ are:
\begin{subequations}
\label{eq:grad_update}
\begin{equation}
    \tau_i^N:=\Pi_\mathbb{T}\Big(\tau_i^N+\alpha(\lambda_\tau^N-\lambda_\tau^S)\Big)
\end{equation}
\begin{equation}
    \rho_i^N:=\Pi_\mathbb{P}\Big(\rho_i^N+\alpha(\lambda_\rho^S-\lambda_\rho^N)\Big)
\end{equation}
\begin{equation}
    \theta_{i}:=\Pi_{\theta_{i}}\Big(\theta_{i}-\alpha(g_{\theta_{i}}^N+g_{\theta_{i}}^S)\Big)
\end{equation}
\end{subequations}
The rest of the steps are as shown in Algorithm \ref{alg:dist}.
\bibliographystyle{IEEEtran}
\balance
\bibliography{ref}

\end{document}